%% file: main.tex
\documentclass{biometrika}

\usepackage{amsmath}
\usepackage{newtxtext}
\usepackage[subscriptcorrection]{newtxmath}
\usepackage{mathrsfs}
\usepackage[hidelinks]{hyperref}
\usepackage[capitalise,nameinlink]{cleveref}
\crefname{section}{\S}{\S\S}\Crefname{section}{Section}{Sections}
\crefname{subsection}{\S}{\S\S}\Crefname{subsection}{Section}{Sections}
\crefname{figure}{Fig.}{Figs.}\Crefname{figure}{Figure}{Figures}
\crefname{assumption}{Assumption}{Assumptions}
\crefname{algo}{Algorithm}{Algorithms}

\graphicspath{{./art/}}

\renewcommand{\top}{{\mathrm{\scriptscriptstyle T}}}
\newcommand{\R}{\mathbb{R}}
\newcommand{\Cx}{\mathbb{C}}
\newcommand{\pr}{P}
\DeclareMathOperator{\tr}{tr}
\DeclareMathOperator{\diag}{diag}
\DeclareMathOperator{\spec}{spec}

\DeclareMathOperator*{\argmin}{arg\,min}
\newcommand{\op}{{\mathrm{op}}}
\newcommand{\Rh}{\hat R}
\newcommand{\Gh}{\hat G}
\newcommand{\Kry}{\mathscr{K}}
\newcommand{\ev}{\Omega}
\newcommand{\cbar}{\bar c}
\newcommand{\Tch}{\mathcal{T}}
\makeatletter
\def\ps@plain{\let\@mkboth\@gobbletwo\def\@oddhead{}\let\@evenhead\@oddhead\def\@oddfoot{\hfil\thepage\hfil}\let\@evenfoot\@oddfoot}
\makeatother

\begin{document}

\markboth{T. Lu, G. J. Goodhill and H. Hu}{Optimal spectral shrinkage via Krylov approximation}

\title{Optimal spectral shrinkage via Krylov approximation}
\author{Tianfeng Lu}
\affil{Department of Electrical \& Systems Engineering, Washington University in St.~Louis,\\ St.~Louis, Missouri, U.S.A. \email{lut@wustl.edu}}

\author{Geoffrey J. Goodhill}
\affil{Departments of Developmental Biology and Neuroscience, Washington University in St.~Louis,\\ St.~Louis, Missouri, U.S.A. \email{g.goodhill@wustl.edu}}

\author{\and Hong Hu}
\affil{Department of Electrical \& Systems Engineering and Department of Statistics \& Data Science,\\ Washington University in St.~Louis, St.~Louis, Missouri, U.S.A. \email{hu.h@wustl.edu}}

\maketitle

\begin{abstract}
Spectral shrinkage encompasses a broad family of estimators for high-dimensional linear regression that regularize estimation and prediction by applying spectral filters to the sample covariance matrix.
Despite its widespread use, achieving oracle-optimal prediction risk under general covariance structures remains challenging, as the oracle filter requires unknown population quantities and has a closed form only for specific covariance models such as isotropic and spiked models.
We propose \textit{risk-targeted Krylov fitting} (RTK), a data-driven spectral estimator that adapts to the unknown covariance structure and attains the oracle risk asymptotically.
Like partial least squares, RTK operates in an $m$-dimensional Krylov space, but it targets the oracle risk directly by minimizing the leave-one-out risk, which we compute exactly without refitting the model $n$ times.
Under general distributions and covariance structures, RTK attains the optimal prediction risk in the Krylov space up to $O(1/n)$.
Furthermore, its risk approaches that of the optimal analytic filter with $m$ of order $\log n$.
In simulations over a wide range of covariance structures, RTK and a truncated variant were robust to the choice of $m$ and consistently among the spectral estimators closest to the oracle.
In applications to single-cell sequencing data and large-scale neural recordings, the proposed methods achieved higher prediction accuracy than competing estimators on most tasks.
\end{abstract}

\begin{keywords}
Chebyshev polynomial; Cross-validation; Proportional asymptotics; Spectral regularization.
\end{keywords}

\section{Introduction}\label{sec:intro}

We study high-dimensional linear regression, where the number of covariates $p$ and the number of observations $n$ are of the same order. The data are $n$ independent observations from the linear model $y=x^\top\beta^\star+\epsilon$, written in the design matrix $X\in\R^{n\times p}$ and the response vector $y\in\R^n$. Here $\beta^\star\in\R^p$ is the true regression coefficient and the noise $\epsilon$ has mean zero given $x$ and variance $\sigma^2$. The performance of an estimator $\beta$ is measured by its prediction risk, the expected squared error at a new observation $(x_0,y_0)$ independent of $(X,y)$,
\begin{equation}\label{eq:risk}
R(\beta)=E((y_0-x_0^\top\beta)^2\mid X,y)=\sigma^2+(\beta-\beta^\star)^\top\Sigma(\beta-\beta^\star),
\end{equation}
where $\Sigma=E(xx^\top)$ is the population covariance matrix.

An important class of estimators, known as spectral estimators, takes the form $\hat\beta_h=h(S_n)s_n$, where $S_n=X^\top X/n$ and $s_n=X^\top y/n$. The spectral filter $h(\cdot)$ is a scalar function applied to the eigenvalues of $S_n$.
Along a sample eigendirection with eigenvalue $\lambda$, such an estimator multiplies the least squares coefficient by the shrinkage factor $\lambda h(\lambda)$ \citep{FrankFriedman1993}, which is below one where the estimator shrinks and above one where it amplifies.
Ridge regression, principal component regression, early-stopped gradient descent and partial least squares (PLS) are all of this form and differ in the shape of the filter.
The risk of a spectral estimator depends critically on the choice of the filter $h$, which governs how the sample eigendirections are weighted.

This raises a natural question: which filter gives the smallest risk, and how can we achieve the smallest risk given a limited amount of data?
In the proportional regime $n,p\to\infty$ with $p/n\to\gamma\in(0,\infty)$, the risk $R(\hat\beta_h)$ is asymptotically deterministic for ridge regression and many other spectral estimators \citep{DobribanRidge2018AOS,HastieMontanariRossetTibshirani2022,green2025}.
Let $\mathcal F$ be the class of fixed filters $h$ for which $R(\hat\beta_h)$ converges in probability to a deterministic limit. The oracle filter is the filter in $\mathcal F$ with the smallest limiting risk,
\begin{equation}\label{eq:oracle}
h^\star=\argmin_{h\in\mathcal F}R_\infty(h),\qquad R(\hat\beta_h)\xrightarrow{p}R_\infty(h),
\end{equation}
and is therefore a fixed function that does not depend on the realization of the data.
Since $\hat\beta_h=h(S_n)s_n$ depends linearly on $h$ and the prediction risk \eqref{eq:risk} is quadratic in $\beta$, the risk $R(\hat\beta_h)$ depends quadratically on $h$, and so does its limit. Finding $h^\star$ therefore amounts to minimizing a quadratic functional of $h$ determined by $\gamma$, $\sigma^2$, the spectral distribution of $\Sigma$ and the distribution of the signal over the eigenvalues of $\Sigma$, as shown by the limits derived for principal component regression \citep{green2025} and spiked covariance \citep{SelfDistillation2026}.
Because $h^\star$ involves the unknown population quantities $\sigma^2$, $\Sigma$ and $\beta^\star$, the oracle estimator $h^\star(S_n)s_n$ is generally unavailable. Moreover, closed forms of $h^\star$ are known only in special cases, namely a ridge filter when the signal is isotropic \citep{DobribanRidge2018AOS} and a rational function when the covariance is spiked \citep{SelfDistillation2026}.
In the proportional regime, the population quantities required to evaluate $h^\star$ are difficult to estimate accurately.
For example, the spectral distribution of $\Sigma$ can be estimated from that of the sample covariance matrix $S_n$ \citep{ElKaroui2008,Kong2017SpectrumEstimation,LWQuEST2017}, but this inverse problem can be ill-posed when the dimension is comparable to the sample size. 

Instead of computing $h^\star$ from estimates of unknown population quantities such as $\Sigma$ and $\sigma^2$, we ask whether a data-driven spectral estimator can attain the oracle risk $R_\infty(h^\star)$ without explicitly estimating these quantities.
We construct such an estimator $\hat\beta_{\hat h}$ from \eqref{eq:oracle} by replacing the limiting risk $R_\infty(h)$ by a risk estimate $\Rh(h)$ and the class $\mathcal F$ by a space of filters $\mathcal F_n$,
\begin{equation}\label{eq:risk-min}
\hat h=\argmin_{h\in\mathcal F_n}\Rh(h).
\end{equation}
Within this framework, we seek $\mathcal F_n$ and $\Rh$ that satisfy two conditions:
\begin{enumerate}[(ii)]\renewcommand\theenumi{\roman{enumi}}\renewcommand\labelenumi{(\theenumi)}%
\item the minimum risk over $\mathcal F_n$ approaches $R_\infty(h^\star)$;
\item $\Rh(h)$ is consistent for $R(\hat\beta_h)$ uniformly over $\mathcal F_n$.
\end{enumerate}
Together, the two conditions ensure that $\hat\beta_{\hat h}$ attains the oracle risk asymptotically.
In general, condition (i) requires $\mathcal F_n$ to grow with $n$, whereas condition (ii) limits how fast it may grow.

To meet conditions (i) and (ii), we minimize the leave-one-out risk over the Krylov space of spectral estimators, a procedure we call risk-targeted Krylov fitting (RTK).
For condition (i), the space $\mathcal F_n$ must be flexible enough to reproduce the shape of the oracle filter, whose shrinkage factor $\lambda h^\star(\lambda)$ can be non-monotone and exceed one \citep{SelfDistillation2026}, as the simulations of \cref{sec:sims} show. 
Optimally tuned ridge regression and principal component regression cannot approximate such an oracle filter, because their shrinkage factors are monotone and at most one.
We take $\mathcal F_n$ to be the space of polynomials of degree less than $m$, where $m$ may grow with $n$, so that the corresponding estimators $\hat\beta_h=h(S_n)s_n$ form the Krylov space
\[
\Kry_m(S_n,s_n)=\mathrm{span}\{s_n,S_ns_n,\ldots,S_n^{m-1}s_n\},
\]
which we abbreviate to $\Kry_m$.
The best estimator in this space attains the smallest risk
\begin{equation}\label{eq:population_risk_Krylov}
\min_{\deg h<m}R(\hat\beta_h)=\min_{\beta\in\Kry_m(S_n,s_n)}R(\beta),
\end{equation}
which approaches the oracle risk as $m$ grows whenever the oracle filter is continuous on the spectrum (\cref{rem:risk-approx}).
For condition (ii), we take $\Rh$ to be the leave-one-out risk. On $\Kry_m$, the leave-one-out predictions are linear in the polynomial coefficients and follow from a single eigendecomposition of the $n\times n$ matrix $XX^\top/n$, allowing RTK to minimize the leave-one-out risk directly over $\Kry_m$.
When a few eigenvalues lie far above the bulk, using a single polynomial for the whole spectrum can lead to poor resolution, as much of the polynomial degree is devoted to a nearly empty spectral gap, where the filter values do not affect the risk. We therefore also introduce a truncated variant of RTK, whose filter fits a polynomial only near the bulk and gives each separated eigenvalue a free shrinkage factor.

Viewing RTK through conditions (i) and (ii) clarifies how it relates to and differs from two established methods: PLS and cross-validation.
PLS was introduced by \citet{Wold1966} as an iterative latent-variable method and developed for chemometric calibration \citep{Wold1984PLS,hoskuldsson1988pls}. It is now widely used for high-dimensional, correlated measurements in chemistry \citep{Wold2001PLS,MouPLSMS2025,XuPLSRMS2025}, genomics \citep{BoulesteixStrimmer2007} and neuroscience \citep{Krishnan2011,Rumyantsev2020,Ebrahimi2022,Helmer2024commbio}.
The $m$-component PLS regression minimizes the empirical risk (the training error) over the same Krylov space as RTK,
\begin{equation}
  \hat\beta^{\rm PLS}_m
  =\argmin_{\beta\in\Kry_m(S_n,s_n)}
    n^{-1}\|y-X\beta\|^2.
\label{eq:pls-krylov-minimization}
\end{equation}
PLS thus meets condition (i) but not condition (ii): the empirical risk it minimizes is not uniformly consistent for the prediction risk \eqref{eq:risk} in the proportional regime (\cref{sec:pls-objective}), so the risk of PLS first decreases in $m$ and then rises toward that of the least squares solution \citep{Kramer2011PLSDOF}.
Conversely, cross-validation methods meet condition (ii) but not condition (i).
Leave-one-out or generalized cross-validation \citep{Allen1974,Stone1974CV,CravenWahba1979,GolubHeathWahba1979} is commonly applied to spectral shrinkage \citep{BauerPereverzevRosasco2007,LoGerfo2008}, where it tunes the parameter of a filter family fixed in advance (\cref{sec:setup}).
These criteria and related risk estimates are consistent in the proportional regime for ridge regression \citep{PatilWeiRinaldoTibshirani2021,RadMaleki2020}, gradient descent paths \citep{PatilWuTibshirani2024,BellecTan2024} and ensembles of penalized estimators \citep{Bellec2025}.
More generally, cross-validation determines the weights of candidate estimators in stacking \citep{Wolpert1992,Breiman1996}, the super learner \citep{vanderLaan2007} and jackknife model averaging \citep{HansenRacine2012}.
In either case, the risk estimate only selects or weights candidate estimators, whose filters are fixed in advance or fitted to the empirical risk as for stacked PLS fits \citep{Kurnia2021}.
Without a closed form of the oracle filter, the resulting estimator cannot in general attain the oracle risk.
RTK resembles stacking in form, but it determines the filter itself by minimizing a uniformly consistent risk estimate, so it can attain the oracle risk as $m$ grows (\cref{rem:stacking}).

Since RTK minimizes the leave-one-out risk over $\Kry_m$ rather than evaluating it at a fixed estimator, it needs $\Rh$ to be consistent uniformly over $\Kry_m$ as in condition (ii), not merely pointwise.
In return, its finite-sample imprecision as a risk estimate \citep{BHT2024CV} matters much less, because a criterion too noisy to estimate the risk can still locate the minimizer of the risk (\cref{rem:relation2CV}).
Our main result is that the leave-one-out risk meets condition (ii): under moment conditions covering Gaussian, sub-Gaussian and bounded elliptical designs, it is consistent at rate $n^{-1/2}$ uniformly over bounded subsets of $\Kry_m$ (\cref{thm:consistency}).
If the basis vectors of $\Kry_m$ are not nearly collinear, RTK therefore attains the risk of the best estimator in $\Kry_m$ up to a term of order $n^{-1}$, whereas PLS does not in general (\cref{cor:oracle,cor:rtk-pls}).
With $m$ of order $\log n$, $\Kry_m$ meets condition (i) whenever the oracle filter is analytic near the spectrum, and the risk of RTK then approaches the oracle risk at a polynomial rate in $n$ (\cref{thm:sieve}).
The only tuning parameter in RTK is $m$. Unlike the level of ridge regression, it does not select a filter but only sets the dimension of $\Kry_m$, within which the leave-one-out risk determines the filter (\cref{rem:stacking}). Each unnecessary dimension costs only $O(n^{-1})$ (\cref{cor:oracle}(b)), so $m$ can be fixed in advance.
The truncated variant has analogous guarantees under a gap condition (\cref{rem:gap}).

In simulations over a wide range of covariance structures and signals (\cref{sec:sims}), RTK and its truncated variant were more accurate than PLS in every setting, by up to 63\%, and unlike that of PLS, the risk of RTK rose only mildly as $m$ grew (\cref{fig:pls-vs-rtk}).
The truncated variant was the most accurate among the spectral estimators compared in more than half of the settings. In the other settings its gap to the best estimator shrank as $n$ grew, also where a competing filter family contains the closed-form oracle. At $n=1000$ its excess risk was never more than about 1\% above the best (\cref{sec:sim-results}).
In single-cell sequencing data and large-scale neural recordings, RTK and its truncated variant improved prediction over the competing spectral estimators in most tasks (\cref{sec:application}).

\section{Spectral shrinkage and partial least squares}\label{sec:pls}

\subsection{Model and spectral estimators}\label{sec:setup}

Formally, the observations $(x_i,y_i)$ $(i=1,\ldots,n)$ are independent copies of $(x,y)\in\R^p\times\R$ following the linear model
\begin{equation}\label{eq:model}
y=x^\top\beta^\star+\epsilon,\qquad E(x)=0,\quad E(xx^\top)=\Sigma,\quad E(\epsilon\mid x)=0,\quad E(\epsilon^2)=\sigma^2 ,
\end{equation}
where the rows of $X$ are $x_i^\top$, the entries of $y$ are $y_i$, and $S_n=X^\top X/n$ and $s_n=X^\top y/n$ are the sample counterparts of $\Sigma$ and $\Sigma\beta^\star$.

With the eigendecomposition $S_n=\sum_{j=1}^p\hat\lambda_j\hat v_j\hat v_j^\top$ and $\hat\lambda_1\geq\cdots\geq\hat\lambda_p\geq0$, a spectral estimator is
\[
\hat\beta_h=h(S_n)s_n=\sum_{j=1}^ph(\hat\lambda_j)(\hat v_j^\top s_n)\hat v_j .
\]
For $\hat\lambda_j>0$, its coefficient along $\hat v_j$ is the least squares coefficient $\hat v_j^\top s_n/\hat\lambda_j$ multiplied by the shrinkage factor $g_j=\hat\lambda_jh(\hat\lambda_j)$.
Ridge regression with level $\mu>0$, principal component regression with threshold $\tau$ and $t$ steps of gradient descent with step size $\eta$ have the filters
\[
h_{\rm ridge}(\lambda)=\frac{1}{\lambda+\mu},\qquad h_{\rm PCR}(\lambda)=\frac{1\{\lambda>\tau\}}{\lambda},\qquad h_{\rm GD}(\lambda)=\frac{1-(1-\eta\lambda)^t}{\lambda}.
\]
Their shrinkage factors are non-decreasing in $\lambda$ and at most one, for gradient descent when $\eta\hat\lambda_1\leq1$. PLS uses a polynomial filter of degree less than $m$ whose coefficients are fitted by least squares (\cref{sec:pls-objective}).

\subsection{Partial least squares and the Krylov space}\label{sec:pls-objective}

We briefly review PLS to define the quantities in \cref{sec:loo} and to motivate risk-targeted Krylov fitting.
By \eqref{eq:pls-krylov-minimization}, PLS minimizes the empirical risk over the Krylov space $\Kry_m(S_n,s_n)=\{q(S_n)s_n:\deg q<m\}$, which we abbreviate to $\Kry_m$. For a univariate response, this characterization and the equivalence of PLS with the $m$-th iterate of conjugate gradients for $S_n\beta=s_n$ started at zero are classical \citep{Wold1984PLS,Helland1988,Phatak2002PLSCGA}.

To parametrize $\Kry_m$, we pick a basis $\varphi_1,\ldots,\varphi_m$ of the polynomials of degree less than $m$, for instance the monomials $\varphi_k(\lambda)=\lambda^{k-1}$ or the Chebyshev polynomials $\varphi_k(\lambda)=\Tch_{k-1}(2\lambda/b-1)$ on an interval $[0,b]$ that contains the spectrum. The choice of basis affects only numerical stability. \Cref{thm:local} and \cref{alg:loo} are stated in the monomial basis for clarity, whereas our implementation uses the Chebyshev basis (\cref{rem:impl}).

Let $b_k=\varphi_k(S_n)s_n$ and $B=(b_1,\ldots,b_m)\in\R^{p\times m}$. Every $\beta\in\Kry_m$ is then $\beta=Bc$ for a coefficient vector $c\in\R^m$. With the in-sample features $V=XB$, that is, $V_{ik}=x_i^\top b_k$ $(i=1,\ldots,n;\ k=1,\ldots,m)$, the empirical risk in \eqref{eq:pls-krylov-minimization} is the least squares criterion
\[
n^{-1}\|y-Vc\|^2=n^{-1}\|y\|^2-2c^\top B^\top s_n+c^\top B^\top S_nBc,
\]
so that $\hat\beta^{\rm PLS}_m=B\hat c^{\rm PLS}_m$, where $\hat c^{\rm PLS}_m$ solves its normal equations
\begin{equation}\label{eq:pls-normal}
V^\top Vc=V^\top y.
\end{equation}

However, our goal is to attain the smallest risk \eqref{eq:population_risk_Krylov} in $\Kry_m$. By \eqref{eq:risk}, the risk $R(c)$ of $\beta=Bc$ is also a quadratic function of $c$,
\begin{equation}\label{eq:risk-quad}
R(c)=E((y_0-x_0^\top Bc)^2\mid X,y)=R(0)-2c^\top\xi+c^\top Gc,\quad G=B^\top\Sigma B,\quad\xi=B^\top\Sigma\beta^\star,
\end{equation}
with $R(0)=\sigma^2+\beta^{\star\top}\Sigma\beta^\star$. The empirical risk minimized by PLS has the form of \eqref{eq:risk-quad} with $(B^\top S_nB,B^\top s_n)$ in place of $(G,\xi)$, that is, with $\Sigma$ and $\Sigma\beta^\star$ replaced by their sample counterparts $S_n$ and $s_n$. PLS thus minimizes a plug-in estimate of the risk built from the same $S_n$ and $s_n$ that span $\Kry_m$. The plug-in is accurate in low-dimensional regimes.

In the proportional regime, $B^\top S_nB$ and $B^\top s_n$ do not estimate $G$ and $\xi$ consistently.
PLS therefore searches a rich space, but optimizes a criterion that is not aligned with the target risk. In line with this, the PLS coefficient is inconsistent when $p/n$ does not vanish \citep{Chun2010sparsePLS}, and its prediction rates depend on how the signal is aligned with the covariance \citep{Cook2019PLSprediction}. 
Analyses of conjugate gradients on random covariance matrices describe how fast the iterates approach the least squares solution \citep{Ding2022CGA}, which itself is a poor estimator in high dimensions.
Since PLS is the $m$-th iterate of conjugate gradients, the non-monotone dependence of its risk on $m$ is an instance of the semi-convergence of Krylov methods for ill-posed problems \citep{Hansen2010}. Hybrid methods aim to overcome semi-convergence by Tikhonov regularization of the projected problem at each iteration \citep{KilmerOLeary2001,ChungNagyOLeary2008}.

\section{Risk-targeted Krylov fitting}\label{sec:loo}

This section introduces risk-targeted Krylov fitting, which minimizes the leave-one-out risk over the Krylov space $\Kry_m$ instead of the empirical risk minimized by PLS (\cref{sec:rtk}). The leave-one-out risk consistently estimates the prediction risk over the Krylov space, and its minimization is a least squares problem. The leave-one-out risk is computed without refitting by \cref{thm:local} and \cref{alg:loo} (\cref{sec:loo-features}), and \cref{alg:rtk} summarizes the method. \Cref{sec:trunc} proposes a truncated variant that adapts the polynomial filter to spectra with a few eigenvalues far above the bulk.

\subsection{Leave-one-out risk over the Krylov space}\label{sec:rtk}

Our goal is to find the coefficient vector $c$ that minimizes the prediction risk $R(c)$ in \eqref{eq:risk-quad}, which is equivalent to minimizing the risk over polynomial filters in \eqref{eq:population_risk_Krylov}.
Since $R(c)$ depends on the unknown $\Sigma$ and $\beta^\star$, it cannot be computed directly, and we estimate it from the observations $(x_i,y_i)$ $(i=1,\ldots,n)$ by the leave-one-out risk
\[
\Rh(c)=n^{-1}\sum_{i=1}^n(y_i-x_i^\top\beta^{(i)})^2,\quad\beta^{(i)}=\sum_{k=1}^mc_k\varphi_k(S_n^{(i)})s_n^{(i)},
\]
where $S_n^{(i)}$ and $s_n^{(i)}$ are $S_n$ and $s_n$ computed without observation $i$, with the normalization $1/n$ kept. The leave-one-out fit $\beta^{(i)}$ keeps the coefficients $c$ and rebuilds only the basis without observation $i$, so $\Rh(c)$ is the cross-validation error of the fixed filter $\sum_kc_k\varphi_k$. Since $\beta^{(i)}$ is a function of the observations other than $i$, the conditional expectation of the $i$-th term given these observations is $R(\beta^{(i)})$ by \eqref{eq:risk}, so that $E(\Rh(c))=E(R(\beta^{(1)}))$ for fixed $c$. The leave-one-out risk is thus unbiased for the expected risk of $\beta$ computed from $n-1$ observations.

Risk-targeted Krylov fitting (RTK) chooses the coefficients $c$ by minimizing the leave-one-out risk $\Rh(c)$,
\begin{equation}\label{eq:rtk}
\hat c^{\rm RTK}_m=\argmin_{c\in\R^m}\Rh(c),\quad\qquad\hat\beta^{\rm RTK}_m=B\hat c^{\rm RTK}_m .
\end{equation}
The minimization in \eqref{eq:rtk} is again a least squares problem over $c$. Indeed, the leave-one-out prediction $x_i^\top\beta^{(i)}=\sum_{k=1}^mc_kU_{ik}$ is linear in $c$, where the leave-one-out features
\[
U_{ik}=x_i^\top\varphi_k(S_n^{(i)})s_n^{(i)}\quad(i=1,\ldots,n;\ k=1,\ldots,m)
\]
are the predictions at $x_i$ of the basis of $\Kry_m(S_n^{(i)},s_n^{(i)})$. Hence $\Rh$ is the least squares criterion
\begin{equation}\label{eq:loo-quad}
\Rh(c)=n^{-1}\|y-Uc\|^2=\Rh(0)-2c^\top\hat\xi+c^\top\Gh c,\qquad\Gh=U^\top U/n,\quad\hat\xi=U^\top y/n,
\end{equation}
with $\Rh(0)=\|y\|^2/n$, which has the form of \eqref{eq:risk-quad} with $(\Gh,\hat\xi)$ in place of $(G,\xi)$. Thus, the minimizer $\hat c^{\rm RTK}_m$ in \eqref{eq:rtk} solves the normal equations
\begin{equation}\label{eq:rtk-normal}
U^\top Uc=U^\top y .
\end{equation}
Compared with the normal equations \eqref{eq:pls-normal} of PLS, \eqref{eq:rtk-normal} replaces the in-sample features $V_{ik}=x_i^\top b_k$ by the leave-one-out features $U_{ik}$, while the Krylov basis $B$ is unchanged.

Note that unbiasedness of $\Rh(c)$ holds only for fixed $c$, whereas $\hat c^{\rm RTK}_m$ depends on the data. Minimizing $\Rh$ therefore requires it to be consistent for $R$ uniformly over bounded sets of coefficients, which \cref{sec:consistency} proves. \Cref{sec:consequences} then shows that RTK attains the risk of the best Krylov estimator up to $O(n^{-1})$ whereas PLS does not, and \cref{sec:sieve} shows that with a slowly growing dimension its risk approaches that of the best analytic filter, which is the oracle risk when the oracle filter is analytic. An unnecessary dimension costs only estimation error of order $n^{-1}$ rather than a drift toward interpolation as in PLS, so $m$ need not be tuned and can be fixed in advance, as long as it is small compared with $n$.

\begin{remark}\label{rem:stacking}
We derived RTK from PLS by keeping its Krylov space and replacing its criterion, and analyzed its statistical properties along the same route. RTK also admits a stacking view.
The columns of $U$ are the leave-one-out predictions of $b_1,\ldots,b_m$ and thus form the level-one data of stacking \citep{Wolpert1992,Breiman1996}. Like stacking, RTK regresses $y$ on these predictions by least squares in \eqref{eq:rtk}. Stacking usually constrains the weights to be nonnegative \citep{Breiman1996}, whereas RTK leaves its coefficients unconstrained. This difference reflects two more fundamental distinctions.
First, the candidates of stacking are estimators that predict well on their own. Their filters are fixed in advance or fitted to the empirical risk, as for ridge estimators at given levels or the PLS fits stacked by \citet{Kurnia2021}. Stacking thus extends model selection by cross-validation, and only the stacking weights target the risk.
In RTK, the vectors $b_k$ are not good estimators but a basis of $\Kry_m$, and the spectral filter of RTK directly targets the risk. Since the leave-one-out risk is consistent uniformly over bounded coefficients (\cref{thm:consistency}), the coefficients need no constraints.
Second, the single parameter $m$ sets the size of the nested Krylov space without specifying the form of any candidate filter, which allows $m$ to grow with $n$.
With a slowly growing $m$, the risk of RTK approaches the oracle risk when the oracle filter is analytic (\cref{thm:sieve}).
\end{remark}

\subsection{Efficient computation of the leave-one-out features}\label{sec:loo-features}

By \eqref{eq:rtk}, risk-targeted Krylov fitting requires only the leave-one-out features $U$. Computing $U$ by refitting would require $n$ fits, and we instead obtain it exactly and efficiently from full-sample quantities.
Writing $\varphi_k(\lambda)=\sum_{a=0}^{m-1}\varphi_{k,a}\lambda^a$, each feature is a fixed linear combination $U_{ik}=\sum_{a=0}^{m-1}\varphi_{k,a}u_{i,a+1}$ of the leave-one-out moments $u_{i,a}=x_i^\top(S_n^{(i)})^{a-1}s_n^{(i)}$ $(a=1,\ldots,m)$. Their full-sample counterparts are the local response moments $v_{i,a}=x_i^\top S_n^{a-1}s_n=(T_n^ay)_i$ with $T_n=XX^\top/n$. The following theorem recovers the leave-one-out moments $u_{i,a}$ from two full-sample quantities, the local response moments $v_{i,a}$ and the local spectral moments $(T_n^j)_{ii}$.

\begin{theorem}\label{thm:local}
Let $u_{i,a}=x_i^\top(S_n^{(i)})^{a-1}s_n^{(i)}$ and $v_{i,a}=(T_n^ay)_i$ for $a\geq1$, with $u_{i,0}=v_{i,0}=y_i$. 
Then for $i=1,\ldots,n$ and all $a\geq1$,
\begin{equation}\label{eq:local-rec}
u_{i,a}=v_{i,a}-\sum_{j=1}^a(T_n^j)_{ii}\,u_{i,a-j}.
\end{equation}
\end{theorem}

The proof of \cref{thm:local} is purely algebraic and is given in the \hyperref[app:proofs]{Appendix}.
By \cref{thm:local}, the leave-one-out moments $u_{i,a}$ require only $v_{i,a}$ and $(T_n^j)_{ii}$, and both follow from one eigendecomposition $T_n=Q\diag(\hat\lambda)Q^\top$ as $(T_n^j)_{ii}=\sum_lQ_{il}^2\hat\lambda_l^j$ and $v_{i,a}=\sum_lQ_{il}\hat\lambda_l^a(Q^\top y)_l$. All leave-one-out features therefore cost $O(n^3+n^2m)$ operations, the cost of one singular value decomposition of $X$.

Write $T_n^{(i)}$ and $y^{(i)}$ for $T_n$ and $y$ without the $i$-th row and column or entry, and $t_i$ for the $i$-th column of $T_n$ without its diagonal entry. Then $u_{i,a}=t_i^\top(T_n^{(i)})^{a-1}y^{(i)}$, and the proof uses only this form and the symmetry of $T_n$. The recursion \eqref{eq:local-rec} therefore holds for any symmetric matrix in place of $T_n$, which covers kernel regression with $T_n$ replaced by $K/n$ for a kernel matrix $K$ and leave-one-out predictions along gradient descent paths with $I-\eta T_n$ in place of $T_n$ (\cref*{S-sec:proofs-gd} of the Supplementary Material).

The same form gives an algorithm that requires only matrix-vector products. Deleting observation $i$ amounts to setting the $i$-th entry of a vector to zero before each multiplication by $T_n$, and keeping one column per observation carries out all $n$ deletions at once.

\begin{samepage}\csname @beginparpenalty\endcsname=10000
\begin{algo}[Masked recursion]\label{alg:loo}
The input is $T_n=XX^\top/n$, the response $y$ and the dimension $m$.
\begin{tabbing}
\qquad\enspace Step 1. Set $Y=y\mathbf 1_n^\top$, so that column $i$ of $Y$ is a copy of $y$.\\*
\qquad\enspace Step 2. For $a=1,\ldots,m$, set the diagonal of $Y$ to zero and replace $Y$ by $T_nY$.\\*
\qquad\enspace\phantom{Step 2. }Record $u_{i,a}=Y_{ii}$ for $i=1,\ldots,n$.
\end{tabbing}
\end{algo}
\end{samepage}

By induction, the $i$-th column of $Y$ after iteration $a$ is $(T_nD_i)^ay$ with $D_i=I-e_ie_i^\top$, so its $i$-th entry is the moment $u_{i,a}$ of \cref{thm:local}. Setting the diagonal $Y_{ii}=u_{i,a-1}$ to zero thus removes the self-feedback $\sum_{j=1}^a(T_n^j)_{ii}u_{i,a-j}$ of observation $i$ while the products are formed, in the spirit of the moment estimators of \citet{Kong2017SpectrumEstimation}. \Cref{thm:local} instead lets the self-feedback occur and divides it out afterwards. The algorithm costs $O(mn^3)$ operations, compared with $O(n^3+n^2m)$ through \cref{thm:local}.

With the features of \cref{thm:local}, RTK in \eqref{eq:rtk} can be computed as follows.

\begin{algo}[Risk-targeted Krylov fitting]\label{alg:rtk}
The input is the data $(X,y)$ and the dimension $m$.
\begin{tabbing}
\qquad\enspace Step 1. Form $T_n=XX^\top/n$ and its eigendecomposition $T_n=Q\diag(\hat\lambda)Q^\top$, and set $b=1.01\,\hat\lambda_1$.\\
\qquad\enspace Step 2. Compute the features $U\in\R^{n\times m}$ of $\varphi_k(\lambda)=\Tch_{k-1}(2\lambda/b-1)$ by \cref{thm:local}.\\
\qquad\enspace Step 3. Compute $q_k=\varphi_k(T_n)y$ for $k=1,\ldots,m$ and solve $U^\top Uc=U^\top y$.\\
\qquad\enspace Step 4. Output $\hat\beta=X^\top(c_1q_1+\cdots+c_mq_m)/n$.
\end{tabbing}
\end{algo}

The output equals $Bc$ because $\varphi_k(S_n)X^\top=X^\top\varphi_k(T_n)$.

\begin{remark}\label{rem:impl}
\Cref{thm:local} and \cref{alg:loo} are stated in the monomial basis for clarity. Since the change to the Chebyshev basis is ill-conditioned for large $m$, our implementation works in the Chebyshev basis directly. Both results carry over to this basis and to models with an intercept (\cref*{S-sec:impl-features} of the Supplementary Material).
\end{remark}

\begin{remark}\label{rem:gcv}
\Cref{thm:local} also explains generalized cross-validation and why RTK does not use it. Exact leave-one-out corrects each observation by its own local spectral moments $(T_n^j)_{ii}$. For ridge regression with level $\mu$, the exact leave-one-out residual is $((T_n+\mu I)^{-1}y)_i/((T_n+\mu I)^{-1})_{ii}$, and the local spectral moments enter only through the denominator. Replacing it by its average $\tr((T_n+\mu I)^{-1})/n$ gives the residual of generalized cross-validation \citep{CravenWahba1979,GolubHeathWahba1979}.
Such averaging is accurate only when the $(T_n^j)_{ii}$ concentrate, which fails for high-order moments and under elliptical designs, where $(T_n)_{ii}=\|x_i\|^2/n$ varies with the radius (\cref*{S-tab:local} of the Supplementary Material). Both cases matter for RTK, whose features involve moments up to order $m$ and whose theory covers bounded elliptical designs (\cref{ass:A}). Under an elliptical design, the error of the averaged-diagonal approximation of $\Gh$ did not decrease with $n$ in our simulations, whereas that of exact leave-one-out decreased at rate $n^{-1/2}$ (\cref*{S-fig:consistency} of the Supplementary Material).
Beyond ridge filters, averaging the moments is no longer equivalent to the scalar degrees-of-freedom correction of generalized cross-validation. Along gradient descent paths this correction fails to track the risk \citep{PatilWuTibshirani2024}, whereas the averaged moments did in the Gaussian settings of \cref*{S-fig:gd} of the Supplementary Material.
\end{remark}

\subsection{Truncation to the bulk of the spectrum}\label{sec:trunc}

The dimension $m$ in RTK depends on how well polynomials of degree less than $m$ approximate the oracle filter on the spectrum. 
Polynomials approximate smooth functions well on the compact support of the bulk, but not over a spectrum that spans several scales. 
When the top eigenvalues lie far above the bulk, an interval $[0,b]$ that covers them places the bulk near its endpoint 0, where the approximation is poor. 
In particular, a filter that is singular within distance $\varepsilon$ of an endpoint has polynomial approximation error no smaller than $\exp(-2m(\varepsilon/b)^{1/2})$ (\cref*{S-lem:endpoint} of the Supplementary Material; see \citealp[Ch.~8]{Trefethen2013}), and $\varepsilon/b$ is small when $b$ is set by the outliers. 
With two or more spikes, the rational oracle filter of \citet{SelfDistillation2026} may moreover have a pole between consecutive outliers. 

However, poles in the gaps do not affect the risk, because a spectral estimator depends on $h$ only through its values at the sample eigenvalues, which lie near the limiting spectrum. 
The filter may therefore take any value in the gaps of the spectrum, which contain no eigenvalues asymptotically. Polynomials can still match the oracle filter on the bulk and at the outliers, although the required degree grows with the spread of the spectrum. 
Truncation exploits the same freedom with fewer parameters. It fits the polynomial only on an interval close to the bulk, whose scale is much smaller than that of the whole spectrum. Each of the $K$ eigenvalues far above this interval receives a free shrinkage factor instead. 
The estimator is linear in these factors, so the leave-one-out risk stays quadratic and is minimized by least squares. A rational filter with poles among the outliers would instead require a nonconvex search over the pole locations, typically on a grid of candidates.
On a spectrum of bounded range, truncation thus lowers the degree that is needed rather than the risk that can be attained, and RTK without truncation still approaches the oracle as $n$ grows, with a slightly larger dimension (\cref*{S-sec:exp-truncation} of the Supplementary Material).

Suppose that $\hat\lambda_1>\cdots>\hat\lambda_K>\hat\lambda_{K+1}$, let $P_K=\sum_{j\leq K}\hat v_j\hat v_j^\top$ and $b'=1.01\,\hat\lambda_{K+1}$, and let $\tilde\varphi_k(\lambda)=\Tch_{k-1}(2\lambda/b'-1)$ be the Chebyshev polynomials on the bulk interval $[0,b']$. The truncated Krylov space of dimension $m+K$ consists of the spectral estimators
\begin{equation}\label{eq:trunc}
\beta=\sum_{k=1}^mc_k\tilde\varphi_k(S_n)(I-P_K)s_n+\sum_{j=1}^Kd_j\hat v_j\hat v_j^\top s_n\qquad(c\in\R^m,\ d\in\R^K),
\end{equation}
whose filter is a polynomial of degree less than $m$ on the bulk eigenvalues $\hat\lambda_l$ $(l>K)$ and takes a free value $d_j$ at each of the $K$ separated eigenvalues. The coefficient $d_j$ sets the shrinkage factor $\hat\lambda_jd_j$ of the $j$-th direction.
The leave-one-out version of \eqref{eq:trunc} keeps $(c,d)$ fixed and uses the top $K$ eigenvectors of $S_n^{(i)}$, so its leave-one-out risk is again a least squares criterion in $(c,d)$ with $m+K$ exact leave-one-out features. We call the method risk-targeted Krylov fitting with truncation (RTK-C), which reduces to RTK for $K=0$.
Let $\hat v_j^{(i)}$ be the $j$-th eigenvector of $S_n^{(i)}$ and $P_K^{(i)}=\sum_{j\leq K}\hat v_j^{(i)}\hat v_j^{(i)\top}$. The leave-one-out features of the two kinds of terms form the matrices $\tilde U\in\R^{n\times m}$ and $O\in\R^{n\times K}$ with entries
\[
\tilde U_{ik}=x_i^\top\tilde\varphi_k(S_n^{(i)})(I-P_K^{(i)})s_n^{(i)},\qquad o_{ij}=x_i^\top\hat v_j^{(i)}\hat v_j^{(i)\top}s_n^{(i)}.
\]
RTK-C minimizes $n^{-1}\|y-\tilde Uc-Od\|^2$, so its coefficients in \eqref{eq:trunc} solve the normal equations
\[
\begin{pmatrix}\tilde U^\top\tilde U&\tilde U^\top O\\O^\top\tilde U&O^\top O\end{pmatrix}\begin{pmatrix}c\\d\end{pmatrix}=\begin{pmatrix}\tilde U^\top y\\O^\top y\end{pmatrix}.
\]
The bulk features $\tilde U$ differ from $U$ because they are deflated by the top $K$ leave-one-out eigenvectors, and the features $O$ of the separated directions are new.

Fitting RTK requires only the leave-one-out moments of \cref{thm:local}. The truncated space further requires the leave-one-out eigenvalues and eigenvectors of the $K$ separated directions, which the following proposition obtains from the same full-sample eigendecomposition through \cref{lem:schur} in the Appendix.

\begin{proposition}
  \label{prop:trunc-loo}
  Let $T_n=Q\diag(\hat\lambda)Q^\top$ and $\tilde y=Q^\top y$, suppose $\hat\lambda_1>\cdots>\hat\lambda_K>\hat\lambda_{K+1}$, and suppose that $Q_{il}\neq0$ for $l\leq K+1$, that is, observation $i$ has a nonzero loading on each of the top $K+1$ eigenvectors of $T_n$. For $j\leq K$, the function $\psi_i(\theta)=\sum_lQ_{il}^2/(\hat\lambda_l-\theta)$ has exactly one root $\hat\lambda^{(i)}_j$ in $(\hat\lambda_{j+1},\hat\lambda_j)$, which is the $j$-th largest eigenvalue of $T_n^{(i)}$ and is simple. The feature of the $j$-th separated direction is
  \[
    o_{ij}=\frac{\sum_lQ_{il}\tilde y_l/(\hat\lambda_l-\hat\lambda^{(i)}_j)}{\sum_lQ_{il}^2/(\hat\lambda_l-\hat\lambda^{(i)}_j)^2},
  \]
  and the bulk features are $\tilde U_{ik}=x_i^\top\tilde\varphi_k(S_n^{(i)})s_n^{(i)}-\sum_{j\leq K}\tilde\varphi_k(\hat\lambda^{(i)}_j)o_{ij}$, where the first term is given by \cref{thm:local}.
\end{proposition}

The proof of \cref{prop:trunc-loo} is given in the \hyperref[app:proofs]{Appendix}.
The bulk feature is the Krylov feature from \cref{thm:local} minus the contributions of the $K$ separated leave-one-out directions.
All leave-one-out features of RTK-C therefore still follow from one eigendecomposition used for \cref{thm:local}, and since $\psi_i$ increases between consecutive full-sample eigenvalues, bisection finds the $nK$ roots at cost $O(n^2K)$ per step. \Cref*{S-sec:impl-proj} of the Supplementary Material details the numerically stable implementation for spectra spanning several orders of magnitude.

In practice, we set $K$ to the number of eigenvalues of $T_n$ that exceed $4$ times their 90\% quantile, capped at $8$. This rule, which we call the resolution rule, removes the directions that would stretch the polynomial interval far beyond the bulk and gives $K=0$ when no eigenvalue lies far above it. In the simulations, it separated the settings in which truncation helps (\cref*{S-sec:exp-krule} of the Supplementary Material).
The bulk dimension $m$ is fixed in advance as for RTK in \cref{sec:rtk}.

\section{Theory}
\label{sec:theory}

This section shows that RTK nearly attains the smallest prediction risk in the $m$-dimensional Krylov space and approaches the oracle risk as $m$ grows with $n$.
The gap between the risk of RTK and the oracle risk splits into two parts. The estimation error is the gap between the risk of RTK and the smallest risk \eqref{eq:population_risk_Krylov} in $\Kry_m$. It arises because RTK minimizes a data-driven risk estimate $\Rh$ instead of the unknown prediction risk $R$, and it grows with $m$. The approximation error is the gap between the smallest risk in $\Kry_m$ and the oracle risk $R_\infty(h^\star)$, and it decreases as $m$ grows.
If $\Rh$ is close to $R$ uniformly over bounded coefficients, then minimizing $\Rh$ nearly minimizes $R$. \Cref{sec:consistency} proves that the leave-one-out risk has this property, and \cref{sec:consequences} uses it to bound the estimation error for fixed $m$ and to compare RTK with PLS.
\Cref{sec:sieve} shows that the approximation error decays geometrically in $m$ when the oracle filter is analytic near the spectrum. Balancing the two errors then leads to $m$ of order $\log n$ and a polynomial rate of convergence to the oracle risk.

\subsection{Uniform consistency of the leave-one-out risk}\label{sec:consistency}

Under the following conditions, we prove that the risk estimate $\Rh$ is close to the unknown prediction risk $R$ uniformly over estimators in $\Kry_m$ with bounded coefficients.

\begin{assumption}\label{ass:A}
The observations follow the linear model \eqref{eq:model}, and $n,p\to\infty$ with $p/n\to\gamma\in(0,\infty)$. In addition, (i) $\|\Sigma\|_\op$ is bounded; (ii) $\sup_{a\neq0}E((a^\top x)^8)/(a^\top\Sigma a)^4$ and $E(y^8)$ are bounded; (iii) there are constants $C_1$ and $C_2$ such that $\pr(\|S_n\|_\op>C_1)\leq C_2n^{-2}$ for all $n$.
\end{assumption}

Conditions (ii) and (iii) of \cref{ass:A} hold for Gaussian and sub-Gaussian designs $x=\Sigma^{1/2}z$ and for elliptical designs $x=r\Sigma^{1/2}z$ with a bounded radius $r$. For the former, $\pr(\|S_n\|_\op>C_1)$ is even exponentially small in $n$. 
Elliptical designs with unbounded radius can violate (iii), because $\|S_n\|_\op$ then grows with $\max_ir_i^2$.

The bounds below hold for any basis $\varphi_1,\ldots,\varphi_m$ of the polynomials of degree less than $m$ (\cref{sec:pls-objective}) and depend on it only through the largest value of $|\varphi_k|$ on a fixed contour around the spectrum of $S_n$. 
We work on the event of a bounded sample spectrum, $\ev_n=\{\|S_n\|_\op\leq C_1\}$, whose probability tends to one by \cref{ass:A}(iii).
Fix $\delta\in(0,1]$ and let $\Gamma$ be the positively oriented boundary of the $\delta$-neighborhood of $[0,C_1]$ in $\Cx$.
On $\ev_n$, $\Gamma$ encloses the spectrum of $S_n$ at distance at least $\delta$, so that $\varphi_k(S_n)=-(2\pi\mathrm{i})^{-1}\oint_\Gamma\varphi_k(z)(S_n-zI)^{-1}{\rm d}z$ with $\|(S_n-zI)^{-1}\|_\op\leq\delta^{-1}$ on $\Gamma$.
We therefore measure the basis by $M_\Phi=\max_k\sup_{z\in\Gamma}|\varphi_k(z)|$. Below, $C$ denotes a constant that depends only on $\delta$, $\sup_np/n$ and the constants in \cref{ass:A}, not on $m$ or the basis, so that $m$ can grow with $n$ in \cref{sec:sieve}.

\begin{theorem}\label{thm:consistency}
Let $\Gh=U^\top U/n$, $\hat\xi=U^\top y/n$ and $\Rh(0)=\|y\|^2/n$ be the coefficients of the leave-one-out risk \eqref{eq:loo-quad}, and let $G=B^\top\Sigma B$, $\xi=B^\top\Sigma\beta^\star$ and $R(0)$ be those of the population risk \eqref{eq:risk-quad}. Under \cref{ass:A}, for all $k,l\in\{1,\ldots,m\}$,
\[
E(1\{\ev_n\}|\Gh_{kl}-G_{kl}|)\leq CM_\Phi^2n^{-1/2},\  E(1\{\ev_n\}|\hat\xi_k-\xi_k|)\leq CM_\Phi n^{-1/2},\  E(|\Rh(0)-R(0)|)\leq Cn^{-1/2}.
\]
Consequently, for every $\cbar\geq0$,
\[
E\Big(1\{\ev_n\}\sup_{\|c\|\leq\cbar}|\Rh(c)-R(c)|\Big)\leq C(1+\cbar mM_\Phi)^2n^{-1/2}.
\]
\end{theorem}

The proof of \cref{thm:consistency} is given in \cref*{S-sec:proof-consistency} of the Supplementary Material.
The errors of $\Gh$ and $\hat\xi$ matter since they determine the estimation error of RTK. 
The leave-one-out risk $\Rh$ is the quadratic $R$ with estimated coefficients, so a minimizer $\hat c$ of $\Rh$ over a bounded set $\mathcal C$ satisfies
\begin{equation}\label{eq:basic-ineq}
R(\hat c)-\min_{\mathcal C}R\leq2\sup_{c\in\mathcal C}|c^\top(\Gh-G)c-2c^\top(\hat\xi-\xi)|,
\end{equation}
where the constant terms cancel since they do not depend on $c$.
Since $\hat c$ depends on the data, $\Rh$ needs to be close to $R$ uniformly over $\mathcal C$, not only at a fixed $c$. The right-hand side of \eqref{eq:basic-ineq} depends only on $\Gh-G$ and $\hat\xi-\xi$, which \cref{thm:consistency} bounds.

To bound these errors, the proof uses that each feature $U_{ik}=x_i^\top\varphi_k(S_n^{(i)})s_n^{(i)}$ is a function of the leave-one-out quantities $S_n^{(i)}$ and $s_n^{(i)}$ evaluated at the held-out $x_i$.
By the contour representation above, $\varphi_k(S_n^{(i)})$ is an integral of the resolvent $(S_n^{(i)}-zI)^{-1}$ over $\Gamma$, and removing one observation changes the resolvent by a rank-one update. The classical resolvent techniques of random matrix theory \citep[Ch.~9]{BaiSilverstein2010} therefore control the perturbation of these functions.
Existing consistency results for ridge regression \citep{PatilWeiRinaldoTibshirani2021}, gradient descent paths \citep{PatilWuTibshirani2024} and ensembles \citep{Bellec2025} assume $x=\Sigma^{1/2}z$ with independent coordinates of $z$, under which the quadratic forms $x_i^\top Ax_i/n$ concentrate around $\tr(\Sigma A)/n$. Since the contour representation and the rank-one updates are deterministic, our proof needs only bounded moments of these quadratic forms, so \cref{thm:consistency} also covers elliptical designs, for which this concentration fails.
For gradient descent with stable step sizes summing to $T$, $M_\Phi\leq Te^{\delta T}$, so the theorem covers paths with $T$ of order at most $\log n$.

\subsection{Risk bounds and comparison with partial least squares}\label{sec:consequences}

\Cref{thm:consistency} and inequality \eqref{eq:basic-ineq} bound the estimation error of RTK, and the bound improves when the Krylov space is well conditioned. If $G$ is invertible, the risk \eqref{eq:risk-quad} is minimized over $\Kry_m$ by $c^\ast=G^{-1}\xi$, and every other $\beta=Bc$ in $\Kry_m$ has excess risk over $Bc^\ast$
\begin{equation}\label{eq:krylov-excess}
R(c)-R(c^\ast)=(c-c^\ast)^\top G(c-c^\ast)=(\beta-Bc^\ast)^\top\Sigma(\beta-Bc^\ast).
\end{equation}

\begin{samepage}
\begin{corollary}\label{cor:oracle}
Let $\mathcal C=\{c\in\R^m:\|c\|\leq\cbar\}$. Under \cref{ass:A}, the following hold.\par\nopagebreak

(a) A minimizer $\hat c$ of $\Rh$ over $\mathcal C$ satisfies $R(\hat c)-\min_{\mathcal C}R=O_p((1+\cbar mM_\Phi)^2n^{-1/2})$.\par\nopagebreak

(b) If the Krylov space is well conditioned, i.e. $\lambda_{\min}(G)\geq\kappa>0$ with probability tending to one and $\|c^\ast\|\leq\cbar$, then the coefficients $\hat c^{\rm RTK}_m=\Gh^{-1}\hat\xi$ of \cref{sec:rtk} satisfy $\|\hat c^{\rm RTK}_m-c^\ast\|=O_p(n^{-1/2})$ and $R(\hat c^{\rm RTK}_m)-R(c^\ast)=O_p(n^{-1})$.
\end{corollary}
\end{samepage}

The proof of \cref{cor:oracle} is given in \cref*{S-sec:proofs-theory} of the Supplementary Material.
Part (a) holds without conditions on $G$ and is the form used in \cref{sec:sieve}, where $m$ grows. The ball $\mathcal C$ only makes the uniform bound of \cref{thm:consistency} applicable, and RTK itself is unconstrained. Part (b) is the analogue of least squares on $m$ well-conditioned regressors, which has excess risk of order $m\sigma^2/n$.

\begin{remark}\label{rem:relation2CV}
For ordinary least squares and other linearly invariant estimates, \citet{BHT2024CV} show that cross-validation is asymptotically uncorrelated with the risk given the data, so that it estimates the risk averaged over training data realizations.
Risk-targeted Krylov fitting is not linearly invariant, and the proof of \cref{thm:consistency} treats its leave-one-out fluctuation directly by a leave-two-out argument. In the errors of $\Gh$ and $\hat\xi$, this fluctuation is of order $n^{-1/2}$, whereas the bias from fitting on $n-1$ rather than $n$ observations is of order $n^{-1}$, so the fluctuation determines the rate $n^{-1/2}$ in \cref{thm:consistency}.
The difference from \citet{BHT2024CV} is that the leave-one-out risk is minimized rather than reported. As for M-estimators \citep{Huber1967}, a criterion too noisy to estimate the risk can still locate the minimizer of the risk.
Since the risk is flat at its minimizer $c^\ast$, a fluctuation of order $n^{-1/2}$ moves the minimizer by $O_p(n^{-1/2})$ but raises the risk by only $O_p(n^{-1})$ (\cref{cor:oracle}(b)). This holds whether or not the fluctuation is correlated with the risk given the data.
\end{remark}

The next result compares RTK and PLS in the same Krylov space.

\begin{corollary}\label{cor:rtk-pls}
Under the conditions of \cref{cor:oracle}(b),
\[
R(\hat\beta^{\rm PLS}_m)-R(\hat\beta^{\rm RTK}_m)=(\hat c^{\rm PLS}_m-c^\ast)^\top G(\hat c^{\rm PLS}_m-c^\ast)-O_p(n^{-1}).
\]
\end{corollary}

The proof of \cref{cor:rtk-pls} is given in \cref*{S-sec:proofs-theory} of the Supplementary Material.
By \eqref{eq:krylov-excess}, the excess risk of PLS over $R(c^\ast)$ is exactly $(\hat c^{\rm PLS}_m-c^\ast)^\top G(\hat c^{\rm PLS}_m-c^\ast)$, and that of RTK is $O_p(n^{-1})$ by \cref{cor:oracle}(b). Since $B^\top S_nB$ and $B^\top s_n$ do not estimate $G$ and $\xi$ consistently, $\hat c^{\rm PLS}_m$ does not converge to $c^\ast$ in general. The gap therefore persists as $n$ grows, while the excess risk of RTK decays like $n^{-1}$ (\cref*{S-fig:rate} of the Supplementary Material).

\subsection{Approximation of the oracle filter}
\label{sec:sieve}

\cref{thm:consistency} controls the estimation error over a fixed Krylov dimension. To approximate the oracle filter in general covariance, the dimension needs to grow with $n$. A larger $m$ lowers the approximation error but raises the estimation error, and we quantify both by Bernstein ellipses around an interval that contains the spectrum. Let $b>C_1$ be fixed, so that $[0,b]$ contains the spectrum of $S_n$ on $\ev_n$. For $\rho>1$, let $\mathcal E_\rho$ be the closed Bernstein ellipse of $[0,b]$ centered at $b/2$ with semi-axes $b(\rho+\rho^{-1})/4$ along the real axis and $b(\rho-\rho^{-1})/4$ along the imaginary axis.
Two values of $\rho$ play different roles.
First, choose $\rho_\Gamma>1$ such that $\Gamma$ lies in $\mathcal E_{\rho_\Gamma}$. Since $|\Tch_k(2z/b-1)|\leq\rho_\Gamma^k$ for $z\in\Gamma$, the Chebyshev basis satisfies $M_\Phi\leq\rho_\Gamma^{m-1}$, so $\rho_\Gamma$ controls how fast the estimation error of \cref{thm:consistency} grows with $m$.
Second, $\rho_\ast$ measures the smoothness of the filters we compare with, and it controls how fast the approximation error decays with $m$.

For $\rho_\ast>1$ and $M_\ast>0$, let $\mathcal A(\rho_\ast,M_\ast)$ be the class of filters that extend to analytic functions bounded by $M_\ast$ in the interior of $\mathcal E_{\rho_\ast}$. A larger $\rho_\ast$ means a smoother filter, which polynomials of degree less than $m$ approximate on $[0,b]$ with error $O(\rho_\ast^{-m})$ \citep[Ch.~8]{Trefethen2013}. 
\citet{KRRspectral2026} consider a similar class of analytic filter functions with extra shape constraints, namely shrinkage factors in $[0,1]$ that are non-decreasing and ridge-like, whereas $\mathcal A(\rho_\ast,M_\ast)$ requires analyticity on an ellipse that does not change with $n$.
Let $R^\ast_{\mathcal A}=\inf_{h\in\mathcal A}R(h(S_n)s_n)$ be the risk of the best filter in this class given the data ($S_n$, $s_n$) at hand.

\begin{theorem}\label{thm:sieve}
Let $\hat\beta_m$ be risk-targeted Krylov fitting with the Chebyshev basis on $[0,b]$ and coefficients constrained to $\|c\|\leq2M_\ast(1-\rho_\ast^{-2})^{-1/2}$, and let $m_n=\lfloor\log n/(2\log(\rho_\ast\rho_\Gamma^2))\rfloor$. Under \cref{ass:A},
\[
R(\hat\beta_{m_n})\leq R^\ast_{\mathcal A}+O_p(n^{-\alpha}\log^2n),\qquad\alpha=\frac{\log\rho_\ast}{2\log\rho_\ast+4\log\rho_\Gamma}\in(0,1/2).
\]
\end{theorem}

The proof of \cref{thm:sieve} balances a Chebyshev approximation error against the estimation error of \cref{thm:consistency} and is given in \cref*{S-sec:proofs-theory} of the Supplementary Material.
\Cref{fig:sieve} shows this trade-off in a simulation. Ridge filters with levels bounded away from zero and rational filters with poles outside $[0,b]$ belong to some $\mathcal A(\rho_\ast,M_\ast)$.
In particular, if the oracle filter $h^\star$ of \eqref{eq:oracle} belongs to $\mathcal A(\rho_\ast,M_\ast)$, then $R^\ast_{\mathcal A}\leq R(h^\star(S_n)s_n)$, so the risk of RTK exceeds that of $h^\star(S_n)s_n$ by at most $O_p(n^{-\alpha}\log^2n)$, and $R(h^\star(S_n)s_n)$ converges to the oracle risk $R_\infty(h^\star)$.

\begin{remark}\label{rem:risk-approx}
\Cref{thm:sieve} requires the filter to be analytic around $[0,b]$, but RTK also approaches the risk of filters that are only continuous on the spectrum, although without a rate. This covers the oracle for spiked covariance and other rational filters with poles in the gaps of the spectrum, and the filter $1\{\lambda>\tau\}/\lambda$ of principal component regression with $\tau$ in a gap.
The reason is that a spectral estimator depends on its filter only through the values at the sample eigenvalues. Let $\hat\nu_n=\sum_j(\hat v_j^\top s_n)^2\delta_{\hat\lambda_j}$, so that $\|h(S_n)s_n\|^2=\int h^2{\rm d}\hat\nu_n$. 
By \eqref{eq:risk} and the triangle inequality,
\[
\big|(R(\hat\beta_{h_1})-\sigma^2)^{1/2}-(R(\hat\beta_{h_2})-\sigma^2)^{1/2}\big|\leq\|\Sigma\|_\op^{1/2}\Big(\int(h_1-h_2)^2{\rm d}\hat\nu_n\Big)^{1/2},
\]
so the values of a filter where $\hat\nu_n$ has no mass do not matter. 
Let $N\subseteq[0,b]$ be a closed set that contains the support of $\hat\nu_n$ with probability tending to one. If $h$ is continuous on $N$, the Weierstrass theorem gives for every $\epsilon>0$ a polynomial $p_\epsilon$ with $|p_\epsilon-h|\leq\epsilon$ on $N$. The measure $\hat\nu_n$ has total mass $\|s_n\|^2$, so $\int(p_\epsilon-h)^2{\rm d}\hat\nu_n\leq\epsilon^2\|s_n\|^2$. Since $\|s_n\|=O_p(1)$ and $h$ is bounded on the compact set $N$, $R(p_\epsilon(S_n)s_n)\leq R(h(S_n)s_n)+O_p(\epsilon)$.
With $m$ and the coefficient bound growing slowly in $n$, \cref{cor:oracle}(a) then shows that RTK attains $R_\infty(h)$ up to $o_p(1)$ for such $h\in\mathcal F$, and the oracle risk if $h^\star$ is of this form.
\end{remark}

The rate of \cref{thm:sieve} degrades as $\rho_\ast$ approaches $1$, which happens when the filter has a pole whose distance to $[0,b]$ is small compared with the length of the interval. As discussed in \cref{sec:trunc}, a few outlying eigenvalues stretch $[0,b]$, so even a pole at a moderate distance from the bulk falls into this case. The oracle filter may also have poles in the gaps of the spectrum. Such poles do not affect the risk (\cref{rem:risk-approx}), but they lie inside $[0,b]$ and rule out the rate of \cref{thm:sieve}. The truncated space fits the outlying directions separately and approximates the filter only on the much shorter bulk interval $[0,b']$, which excludes these gaps. Compared with the length of this interval, the poles of the filter are typically far away. Let $\mathcal E'_\rho$ and $\mathcal A'(\rho,M)$ denote $\mathcal E_\rho$ and $\mathcal A(\rho,M)$ for this interval.

\begin{proposition}\label{prop:linearize}
Suppose that the restriction of $h$ to $[0,b']$ belongs to $\mathcal A'(\rho',M)$ for some $\rho'>1$ and $M>0$, and $M_{\rm out}=\max_{j\leq K}|h(\hat\lambda_j)|$ is finite.\par\nopagebreak

(a) (Uniform approximation) There is a filter $h_m$ of the form \eqref{eq:trunc} with $h_m(\hat\lambda_j)=h(\hat\lambda_j)$ $(j\leq K)$ and $\sup_{[0,b']}|h_m-h|\leq2M\rho'^{\,1-m}/(\rho'-1)$.\par\nopagebreak

(b) (Risk approximation) On $\ev_n$, $R(h_m(S_n)s_n)-R(h(S_n)s_n)\leq C(1+M+M_{\rm out})^2\rho'^{-m}(1+\|y\|^2/n)$, where $C$ depends also on $\rho'$.\par\nopagebreak

(c) (Rational filters) The restriction to $[0,b']$ of a rational filter $\sum_jw_j/(\lambda+\mu_j)$ with real poles at distance at least $\varepsilon$ from $[0,b']$ belongs to $\mathcal A'(\rho',M)$ with $\rho'=1+(\varepsilon/b')^{1/2}$ and $M=4\sum_j|w_j|/(3\varepsilon)$, wherever the poles lie relative to $\hat\lambda_1,\ldots,\hat\lambda_K$.
\end{proposition}

The proof of \cref{prop:linearize} is given in \cref*{S-sec:proofs-trunc} of the Supplementary Material.
In part (b), the factor $1+\|y\|^2/n=O_p(1)$ bounds the size of $s_n$, since $\|s_n\|^2\leq C_1\|y\|^2/n$ on $\ev_n$.
The positive poles of the oracle for spiked covariance all lie above the smallest outlying eigenvalue, so by part (c) the degree needed on the bulk is governed by the gap between the bulk and the outliers.

\begin{remark}\label{rem:gap}
The results of \cref{sec:consistency,sec:consequences} extend to the truncated Krylov space with a fixed bulk interval under a gap condition (\cref*{S-sec:proofs-trunc} of the Supplementary Material). The condition requires the spectrum of $S_n$ to split with high probability into the bulk and $K$ eigenvalues in fixed windows away from it, also after one or two observations are removed. The fitted shrinkage factors $\hat\lambda_j\hat d_j$ are then within $O_p(n^{-1/2})$ of the risk-optimal factors when the space is well conditioned, and \cref{prop:linearize} gives the analogue of \cref{thm:sieve} for the class $\mathcal A'(\rho',M)$. The condition fails for near-critical spikes, whose outlying eigenvalues lie within the fluctuation scale of the bulk edge.
\end{remark}

\section{Simulations}\label{sec:sims}
\subsection{Design}\label{sec:sim-setup}

The simulations cover 3 parts. The first two examine the theory in \cref{sec:theory}: \cref{sec:sim-pls} follows PLS and RTK in the same Krylov space as $m$ grows, which isolates the gap of \cref{cor:rtk-pls} and the estimation cost of \cref{cor:oracle}(b), and \cref{sec:sim-sieve} follows the approach of RTK to the oracle filter as $n$ grows, which separates the approximation and estimation errors of \cref{thm:sieve}. 
The third part, in \cref{sec:sim-results}, is a systematic benchmark of RTK and RTK-C against 5 spectral estimators over 23 settings, chosen so that the oracle filter takes qualitatively different shapes.

Data are generated from the linear model \eqref{eq:model} with $x=\Sigma^{1/2}z$, $z\sim N(0,I_p)$, $\epsilon\sim N(0,\sigma^2)$ independent of $x$, $\sigma^2=1$ and $p=\gamma n$. Without loss of generality, we take $\Sigma$ diagonal, so that its eigenvectors are the coordinate vectors $e_k$.
Two further settings in the Supplementary Material instead use the elliptical design $x=r\Sigma^{1/2}z$, with $r^2\in\{0.25,1.75\}$ equally likely and independent of $z$, whose coordinates are dependent although uncorrelated.
The settings differ in the spectrum of $\Sigma$ and in how $\beta^\star$ is distributed over its eigenspaces (\cref{tab:settings}). 
Ridge regression is the exact oracle in the isotropic setting. When one or two spikes carry the signal, the oracle amplifies the leading sample directions. When the largest of several spikes carries no signal, the oracle is non-monotone across the separated eigenvalues, and when the signal lies in one atom of a latent-factor or three-atom spectrum, it has a bump inside the sample spectrum. In the power-law settings the sample eigenvalues span several orders of magnitude.

\begin{table}
\tbl{Simulation settings, with $p=\gamma n$}{%
\begin{tabular}{lllll}
\\
Setting & $\gamma$ & Spectrum of $\Sigma$ & Signal & $\|\beta^\star\|^2$\\[5pt]
Isotropic & 1.5$^*$ & $I_p$ & $N(0,2p^{-1}I_p)$ & 2\\
One spike & 1.5 & $\diag(3,1,\ldots,1)$ & $e_1$ & 1\\
Spike + bulk & 2$^*$ & $\diag(5,1,\ldots,1)$ & $1.7e_1$ + bulk direction & 4\\
Two spikes & 1.5 & $\diag(6,3,1,\ldots,1)$ & $0.7e_1+0.8e_2$ + bulk direction & 1.38\\
Multiple spikes & 1.5 & $\diag(32,16,8,4,1,\ldots,1)$ & $e_2+e_3$ & 2\\
Latent factor & 1.5$^*$ & 5\% at 4, 95\% at 1 & in the eigenvalue-4 space & 1.5\\
Three-atom & 1.5 & 5\% at 12, 15\% at 4, 80\% at 1 & in the eigenvalue-4 space & 4\\
Mixed & 1.5 & half 1, half 2, one at 12 & $0.6e_1$ + eigenvalue-2 space & 1.86\\
Power law, $a=0.6$ & 1.5$^*$ & $\lambda_k\propto k^{-0.7}$, mean 1 & $\beta^\star_k\propto\pm k^{-0.6}$ & 2\\
Power law, $a=1$ & 1.5 & $\lambda_k\propto k^{-0.7}$, mean 1 & $\beta^\star_k\propto\pm k^{-1}$ & 2\\
Steep power law & 1.5 & $\lambda_k\propto k^{-0.9}$, mean 1 & $\beta^\star_k\propto\pm k^{-0.6}$ & 2
\end{tabular}}
\label{tab:settings}
\begin{tabnote}
$\lambda_k$, $k$th eigenvalue of $\Sigma$; $a$, decay exponent of the signal. Components in an eigenspace, including the bulk, point in a uniformly random direction within it, scaled so that $\|\beta^\star\|^2$ is as given, and signs in the power-law settings are random. For the isotropic setting $\|\beta^\star\|^2$ is the expectation. In the three-atom setting the proportions are chosen so that the middle atom stays separated from the other two in the sample spectrum. $^*$Also run with $\gamma=3$ (isotropic), $\gamma=1$ (spike + bulk), $\gamma\in\{0.5,2.5\}$ (latent factor) and $\gamma=0.6$ (power law). Seven further settings are given in \cref*{S-tab:settings-more} of the Supplementary Material.
\end{tabnote}
\end{table}

We compare 7 estimators in the benchmark. 
Ridge regression (Ridge) chooses its regularization level by exact leave-one-out; principal component regression (PCR) chooses the number of components by tenfold cross-validation; PLS chooses $m\leq6$ by approximate leave-one-out, which combines its normal equations \eqref{eq:pls-normal} with the leave-one-out features of RTK; gradient descent (GD) chooses its stopping time by exact leave-one-out, computed from \cref{thm:local}; and one-step self-distillation (SD) \citep{DangPatilRinaldo2026} chooses its level and mixing weight by exact leave-one-out. 
Every tuning parameter thus targets the prediction risk, through its cross-validation estimate.
RTK and RTK-C use $m=5$ without further selection, and RTK-C chooses $K$ by the resolution rule of \cref{sec:trunc}. \Cref{sec:sim-sieve,sec:sim-results} examine the choice of $m$. 
\Cref{sec:sim-pls} also includes the hybrid method of \citet{ChungNagyOLeary2008}, whose regularization parameter is chosen at every step by weighted generalized cross-validation. 
As an infeasible benchmark, the oracle filter $h^\star$ of \eqref{eq:oracle} is approximated by the best linear combination of polynomials and resolvents, fitted to $50$ independent draws of each setting at the same $n$. 
For Gaussian designs, whose limiting risk $R_\infty$ is well defined \citep{green2025}, we directly computed $h^\star$ by minimizing the explicit form of $R_\infty(h)$ over the values of $h$ at the points of a fine grid on the limiting spectrum. The two filters nearly coincide, and their asymptotic excess risks differ by at most 0.6\%, so the comparisons with the oracle below do not depend on how the oracle is computed (\cref*{S-sec:exp-rmt-oracle} of the Supplementary Material). Tuning grids and the construction of the oracle are described in \cref*{S-sec:exp-tuning} of the Supplementary Material. We report the excess risk $R(\hat\beta)-\sigma^2$ of \eqref{eq:risk}, computed exactly from $\Sigma$ and $\beta^\star$.

\subsection{Stability in the Krylov dimension}\label{sec:sim-pls}

\begin{figure}
\includegraphics[width=\textwidth]{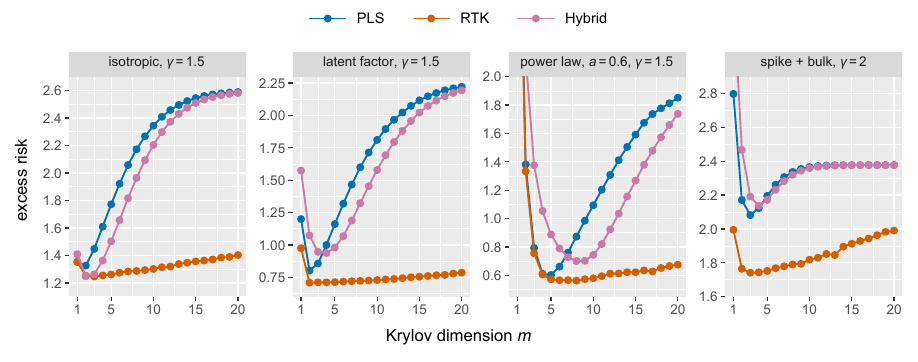}
\setlength{\figwidth}{\textwidth}
\caption{Excess risk against the Krylov dimension $m$ for estimators sharing the same Krylov space: partial least squares minimizing the empirical risk (PLS); risk-targeted Krylov fitting minimizing the exact leave-one-out risk (RTK); and the hybrid method (Hybrid) of \citet{ChungNagyOLeary2008}, which applies Tikhonov regularization to the projected problem with its parameter chosen at every $m$ by weighted generalized cross-validation.
Means over 20 replicates with $n=400$ and the same data for all methods. The settings are described in \cref{tab:settings}, and the values are listed in \cref*{S-tab:A-full} of the Supplementary Material.}
\label{fig:pls-vs-rtk}
\end{figure}

As shown in \cref{fig:pls-vs-rtk}, the risk of PLS was smallest at 2--5 components and then rose toward that of the least squares solution.
The risk of RTK rose by at most 20\% up to $m=20$, consistent with an estimation cost of order $n^{-1}$ per dimension (\cref{cor:oracle}(b)).
At the best number of components of PLS the gap of \cref{cor:rtk-pls} remained: RTK was 5 to 16\% better. The largest gain was in the spike + bulk setting, in which PLS deteriorates little with $m$ but is worse at every $m$.
The hybrid method deteriorated almost as much as PLS, because its regularization parameter decreased with $m$, so that its iterates drifted toward the unregularized conjugate gradient solution.

\subsection{Approach to the oracle}\label{sec:sim-sieve}

\begin{figure}
\includegraphics[width=\textwidth]{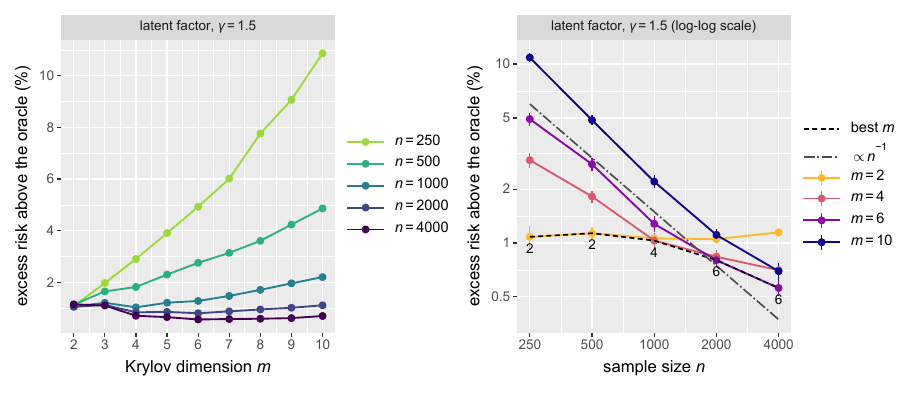}
\setlength{\figwidth}{\textwidth}
\caption{Excess risk of risk-targeted Krylov fitting above that of the oracle filter $h^\star$, as a percentage of the excess risk of the oracle, in the latent-factor setting with $\gamma=1.5$: against the Krylov dimension $m$ for $n$ from 250 to 4000 (left), and against $n$ for fixed $m$ (right). In the right panel the dashed line follows the best $m$ at each $n$ (printed below it), and the dash-dotted line is proportional to $n^{-1}$. Means over 200 ($n\leq500$), 100 ($n=1000$, 2000) and 50 ($n=4000$) replicates. Bars in the right panel are $\pm1$ standard error of the paired differences. Values for $m=1$, about 40\% above the oracle, are not shown.}
\label{fig:sieve}
\end{figure}

In \cref{fig:sieve}, the gap between RTK and the oracle at fixed $m$ is the sum of an approximation error and an estimation error. The first is the excess risk of the best polynomial filter of degree less than $m$ over the oracle and does not depend on $n$, and the second is of order $n^{-1}$ by \cref{cor:oracle}(b).
For each $m$, we regressed the gap across the sample sizes of \cref{fig:sieve} on $n^{-1}$ by weighted least squares, so that the intercept estimates the approximation error and the slope divided by $n$ estimates the estimation error. The residuals were within the Monte Carlo error (\cref*{S-tab:sievefit} of the Supplementary Material), so these two errors account for the gap.
The approximation error decreased roughly geometrically in $m$ and was indistinguishable from zero for $m\geq9$, as expected for an oracle that is analytic near the spectrum, whereas each additional dimension added about $2\sigma^2/n$ to the excess risk, of the same order as the cost $\sigma^2/n$ of a regressor in least squares. In the right panel, the curve for $m=2$ therefore stays at its approximation error for every $n$, whereas that for $m=10$ falls like $n^{-1}$. As the curves cross, the best dimension grows with $n$, from 2 for $n\leq500$ to 6 for $n\geq2000$. Along the best dimension (dashed line), the gap decreased like $n^{-0.44}$ between $n=1000$ and 4000. This is slower than $n^{-1}$, as the exponent $\alpha<1/2$ of \cref{thm:sieve} allows.

Because an unnecessary dimension costs only estimation error, overestimating the dimension is cheaper than underestimating it (\cref{sec:rtk}), and the left panel shows the risk becoming flat in $m$ over a widening range as $n$ grows. The best dimension also exceeded the schedule $m_n$ of \cref{thm:sieve}, which stays at 2--3 over this range for typical constants because its bound allows the cost per dimension to grow exponentially in $m$. In these simulations the cost did not grow. The fixed dimension $m=5$ used in \cref{sec:sim-results} was within 0.2 percentage points of the best one for $n\geq1000$.

\subsection{Comparison with other spectral estimators}\label{sec:sim-results}

\begin{table}
\tbl{Simulations: mean excess risk $R(\hat\beta)-\sigma^2$ over 30 replicates, $n=1000$}{%
\begin{tabular}{lccccccccc}
\\
Setting & $\gamma$ & Ridge & PCR & PLS & GD & SD & RTK & RTK-C & Oracle\\[5pt]
Isotropic & 1.5 & \textbf{1.265} & 1.558 & 1.381 & 1.288 & 1.266 & 1.274 & 1.274 & 1.258\\
One spike & 1.5 & 0.766 & 0.734 & 0.659 & 0.639 & \textbf{0.544} & 0.551 & 0.551 & 0.543\\
Spike + bulk & 2 & 2.173 & 2.211 & 2.071 & 2.029 & 1.687 & \textbf{1.682} & \textbf{1.682} & 1.668\\
Two spikes & 1.5 & 0.872 & 0.874 & 0.719 & 0.744 & 0.664 & \textbf{0.643} & \textbf{0.643} & 0.633\\
Multiple spikes & 1.5 & 0.747 & 0.580 & 0.491 & 0.507 & 0.377 & \textbf{0.299} & 0.302 & 0.293\\
Latent factor & 1.5 & 0.970 & 0.924 & 0.785 & 0.814 & 0.722 & \textbf{0.701} & \textbf{0.701} & 0.691\\
Three-atom & 1.5 & 2.316 & 2.357 & 2.115 & 2.130 & 2.014 & \textbf{1.998} & \textbf{1.998} & 1.945\\
Mixed & 1.5 & 1.522 & 1.730 & 1.525 & 1.477 & \textbf{1.454} & 1.458 & 1.458 & 1.444\\
Power law, $a=0.6$ & 1.5 & 0.628 & 0.593 & 0.663 & 0.515 & 0.502 & 0.558 & \textbf{0.475} & 0.471\\
Power law, $a=1$ & 1.5 & 0.389 & 0.235 & 0.228 & 0.221 & 0.205 & 0.174 & \textbf{0.162} & 0.158\\
Steep power law & 1.5 & 0.468 & 0.441 & 0.990 & 0.387 & 0.388 & 0.860 & \textbf{0.370} & --
\end{tabular}}
\label{tab:main}
\begin{tabnote}
Ridge, ridge regression; PCR, principal component regression; PLS, partial least squares with $m\leq6$ chosen by approximate leave-one-out; GD, gradient descent; SD, one-step self-distillation; RTK, risk-targeted Krylov fitting with $m=5$; RTK-C, risk-targeted Krylov fitting with truncation and $m=5$; Oracle, oracle filter $h^\star$, which is infeasible; --, not available (\cref*{S-sec:exp-tuning} of the Supplementary Material); $a$, decay exponent of the signal. The smallest excess risk among the 7 estimators is in bold. Monte Carlo standard errors of the means are at most 0.038; differences between estimators, which are fitted to the same replicates, are estimated more precisely. Eleven of the 23 settings are shown. All are in \cref*{S-tab:all1000} of the Supplementary Material.
\end{tabnote}
\end{table}

\begin{figure}
\includegraphics[width=\textwidth]{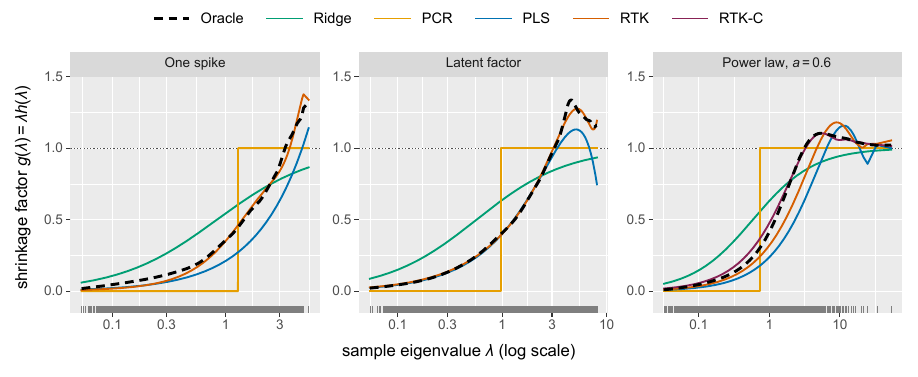}
\setlength{\figwidth}{\textwidth}
\caption{Shrinkage factors $g(\lambda)=\lambda h(\lambda)$ at the sample eigenvalues for the first replicate of the one-spike, latent-factor and power-law ($a=0.6$) settings of \cref{tab:main}, against those of the oracle filter $h^\star$ (dashed). The ticks at the bottom mark the sample eigenvalues, and the dotted line marks $g=1$. In the one-spike and latent-factor settings the resolution rule selects $K=0$, so RTK-C coincides with RTK and is omitted for clarity. Abbreviations as in \cref{tab:main}.}
\label{fig:shrinkage}
\end{figure}

RTK and RTK-C had lower excess risk than PLS in every setting, by up to 63\%. No estimator dominated, and each was best where the oracle has a shape it can express: ridge regression in the isotropic settings and self-distillation in the settings in which a spike carries most of the signal. A competitor whose filter family contains the oracle has no approximation error and estimates one or two parameters instead of $m$, so RTK cannot be expected to improve on it there. Its disadvantage in these settings is mainly an estimation cost of order $n^{-1}$ and shrank from $n=400$ to $n=1000$ (\cref*{S-tab:all400} of the Supplementary Material). The truncated variant came closest to being uniformly good. It had the smallest excess risk in 13 of the 23 settings, and its excess risk was never more than about 1\% above that of the best of the 7 estimators (\cref*{S-tab:regret} of the Supplementary Material).

Its largest gains came where the competing filters cannot follow the oracle: it was 20\% better than self-distillation in the multiple-spike setting and up to 21\% better in the power-law settings, and these gains persisted at $n=400$. These are the only settings in which the resolution rule selected $K>0$, and truncation lowered the excess risk in the power laws by up to 57\%. In the other settings RTK and RTK-C coincide. Choosing $m$ by cross-validation instead of fixing $m=5$ changed the excess risk only slightly (\cref*{S-tab:mdim} of the Supplementary Material).

\Cref{fig:shrinkage} shows how these differences arise. In the one-spike and latent-factor settings the oracle shrinkage factors reach about 1.3 at the top of the spectrum, which ridge regression and principal component regression cannot attain. PLS departs from the oracle there, whereas RTK follows it closely. In the power-law setting the sample eigenvalues span 3 orders of magnitude. As anticipated in \cref{sec:trunc}, RTK then resolves the rise of the oracle on the bulk poorly, whereas free factors at the largest eigenvalues let the truncated variant follow it.

\section{Application to real data}\label{sec:application}

We applied the estimators to single-cell sequencing data and large-scale neural population recordings, in both of which a response is predicted linearly from hundreds to thousands of correlated measurements.
The single-cell data are 8 public data sets (\cref*{S-tab:singlecell} of the Supplementary Material): 4 from 10x Genomics and 4 Perturb-seq screens \citep{Norman2019,Replogle2022,Papalexi2021}. Three of the 10x data sets are of peripheral blood mononuclear cells (PBMC) and the fourth is of embryonic mouse brain. One PBMC data set also includes 17 surface proteins measured by CITE-seq \citep{Stoeckius2017}. After pretreatment as in Seurat \citep{Satija2015} and Scanpy \citep{Wolf2018}, each retains at most 2000 highly variable genes, and each task predicts a marker or perturbation-responsive gene from the other genes, or a surface protein from all genes. The neural data are from the visual cortex. In anesthetized macaques \citep{ZandvakiliKohn2015}, the common fluctuation of the primary visual cortex (V1) population is first regressed out, and the mean activity of the recorded secondary visual cortex (V2) neurons is then predicted from about 100 V1 neurons, as in \citet{Semedo2019}. In mice \citep{Stringer2021}, the angle of a static grating is decoded from 2000 V1 neurons.

All methods use the same random splits into training and test sets and the tuning of \cref{sec:sims}. The methods are compared by the test mean squared error or the out-of-sample $R^2$. Following \cref{sec:sim-sieve}, the Krylov dimension was fixed in advance and increased with the training size: $m=2$ for the macaque data (at most 320 training trials), $m=3$ for the single-cell data (about 1000 training cells) and $m=5$ for the mouse data (2000--4000 training trials). Details are given in \crefrange*{S-sec:exp-singlecell}{S-sec:exp-mouse} of the Supplementary Material.

\begin{table}
\tbl{Single-cell sequencing: test error relative to ridge regression (smaller is better)}{%
\begin{tabular}{llccccccc}
\\
Data set & Targets & PCR & PLS & GD & SD & RTK & RTK-C & W/T/L\\[5pt]
PBMC 3k & 9 genes & 0.954 & 0.961 & 0.947 & 0.950 & 0.939 & \textbf{0.934} & 6/1/2\\
PBMC 10k & 13 genes & 0.978 & 0.991 & 0.972 & 0.971 & 1.031 & \textbf{0.956} & 10/3/0\\
Mouse brain 10k & 11 genes & 0.970 & 0.991 & 0.973 & 0.976 & 1.002 & \textbf{0.966} & 5/6/0\\
CITE-seq & 14 proteins & 0.981 & 0.987 & 0.972 & 0.970 & 1.013 & \textbf{0.950} & 11/3/0\\[3pt]
Norman 2019, K562 & 10 genes & 0.994 & 0.985 & 0.973 & 0.966 & 0.963 & \textbf{0.958} & 6/2/2\\
Replogle 2022, K562 & 10 genes & 0.987 & 1.003 & 0.975 & 0.975 & 0.970 & \textbf{0.963} & 7/3/0\\
Replogle 2022, RPE1 & 10 genes & 0.975 & 1.003 & 0.978 & 0.982 & 0.976 & \textbf{0.963} & 8/2/0\\
Papalexi 2021, THP-1 & 10 genes & 0.983 & 1.002 & 0.972 & 0.973 & 0.989 & \textbf{0.961} & 9/1/0\\
Papalexi 2021, THP-1 & 4 proteins & 1.003 & 1.033 & \textbf{0.992} & 0.995 & 0.993 & 1.003 & 0/1/3
\end{tabular}}
\label{tab:singlecell-results}
\begin{tabnote}
Abbreviations as in \cref{tab:main}; W/T/L, numbers of targets in which RTK-C was better than, indistinguishable from or worse than the best of the 5 competing estimators for that target, judged by a paired $t$-statistic over the splits beyond $\pm2$; $p$, number of predictors; $n_{\rm train}$, number of training cells. Entries are medians over the targets of the ratio of the test mean squared error to that of ridge regression, averaged over 12 random splits with $n_{\rm train}=p/1.5$, and RTK and RTK-C use $m=3$. Medians are reported because the targets differ widely in how well they can be predicted. The smallest median among the 7 estimators is in bold.
\end{tabnote}
\end{table}

\begin{table}
\tbl{Neural recordings: mean out-of-sample $R^2$ (larger is better)}{%
\begin{tabular}{crcccccccc}
\\
$p/n_{\rm train}$ & $n_{\rm train}$ & $m$ & Ridge & PCR & PLS & GD & SD & RTK & RTK-C\\[5pt]
\multicolumn{10}{l}{Macaque V1 to V2: prediction of the averaged V2 activity}\\
1.5 & 59--105 & 2 & 0.464 & 0.439 & 0.451 & 0.469 & 0.448 & \textbf{0.470} & 0.469\\
1 & 88--158 & 2 & 0.497 & 0.480 & 0.484 & 0.505 & 0.483 & \textbf{0.506} & 0.503\\
0.5 & 176--316 & 2 & 0.536 & 0.519 & 0.525 & \textbf{0.543} & 0.536 & \textbf{0.543} & 0.538\\
0.28--0.49 & 320 & 2 & 0.545 & 0.528 & 0.539 & \textbf{0.550} & 0.547 & \textbf{0.550} & 0.548\\[3pt]
\multicolumn{10}{l}{Mouse V1: decoding of the grating angle}\\
0.53--0.57 & 3490--3759 & 5 & 0.619 & 0.604 & 0.616 & \textbf{0.629} & 0.628 & 0.626 & \textbf{0.629}\\
0.85--0.92 & 2181--2349 & 5 & 0.590 & 0.579 & 0.589 & \textbf{0.603} & \textbf{0.603} & 0.602 & \textbf{0.603}
\end{tabular}}
\label{tab:neuro}
\begin{tabnote}
Abbreviations as in \cref{tab:main}; $p$, number of neurons; $n_{\rm train}$, number of training trials; $m$, Krylov dimension of RTK and RTK-C. The macaque data comprise 40 data sets with 88 to 158 V1 neurons and 400 trials, with 100 random splits each, and the mouse data 12 tasks with 2000 neurons and 4362 to 4699 trials, with 3 random splits each. Means are reported because each row predicts the same type of response in all data sets or tasks. The largest $R^2$ among the 7 estimators is in bold.
\end{tabnote}
\end{table}

In the single-cell data the truncated variant was most accurate in all groups of targets but one, and it was better than the best competing estimator in most tasks (\cref{tab:singlecell-results}). The exception is the surface proteins of the THP-1 screen, which all methods predict poorly. 
Without truncation, RTK lost mainly in the 10x data sets with the widest spectra, where the leading eigenvalues stretch the interval on which the polynomial needs to be accurate, as the argument of \cref{sec:trunc} predicts. 
In the neural data the methods were closer to one another (\cref{tab:neuro}). RTK or RTK-C had the largest or nearly the largest out-of-sample $R^2$ at every training size. 
RTK was best in the macaque data, in which only 2--3 eigenvalues were separated from the bulk, and early-stopped gradient descent and self-distillation came close at the larger training sizes. 
The leading eigenvalues of the mouse data lie far outside the bulk, and there RTK-C, early-stopped gradient descent and self-distillation had mean $R^2$ within 0.001 of one another. 
Ridge regression, principal component regression and PLS were less accurate than RTK-C in most groups of single-cell targets and than both RTK and RTK-C at every training size in the neural data.

\section{Discussion}
\label{sec:discussion}

We proposed risk-targeted Krylov fitting, which targets the oracle risk directly instead of estimating the oracle filter. The oracle risk, previously attained only by filters tailored to special covariance structures, can thus be approached adaptively, without prior knowledge of the structure.
More broadly, RTK follows two principles: nested spaces of spectral estimators whose smallest risk tends to the oracle one, and a criterion that estimates the risk uniformly over these spaces. 
This view explains why the Krylov space of PLS is a good starting point, even though the empirical risk that PLS minimizes is not.
First, the shrinkage factors of PLS oscillate and can exceed one \citep{FrankFriedman1993,Shrinkagestructure2000,peculiarPLS2000,Kramer2007}. These long-recognized peculiarities show that the Krylov space contains filters of the shapes that the oracle filter may require, whereas the more common spectral estimators shrink monotonically and never amplify. 
Second, the leave-one-out features of the Krylov space follow from one eigendecomposition through the recursion of \cref{thm:local}, so that the exact leave-one-out risk can be minimized without refitting.

This view also suggests directions for future work, mainly in the choice of the space. Truncation is one such choice, and it helps only when a few eigenvalues lie far above the bulk. On spectra spanning a wide range of scales, rational approximation with geometrically spaced poles typically outperforms polynomials \citep{RationalKrylov2013}, which suggests rational Krylov spaces \citep{RationalKrylov1998} spanned by the ridge estimators $(S_n+\mu_kI)^{-1}s_n$ at fixed $\mu_k>0$. Their exact leave-one-out features follow from \cref{lem:schur} with $z=-\mu_k$, and the proof of \cref{thm:consistency} carries over because these filters are holomorphic near the spectrum. What remains is the choice of the poles and an analogue of \cref{thm:sieve}. 
Beyond linear regression, the leave-one-out features of \cref{thm:local} are exact for kernel regression in any regime, and the theory of \cref{sec:theory} may extend to inner-product kernels in the polynomial regime $n\asymp p^{\zeta}$ with $\zeta>0$ \citep{HuLuMisiakiewicz2024,KRRspectral2026}. Block Krylov spaces suggest a further extension to multivariate responses \citep{Cook2013EnvelopePLS}. 
Finally, any estimate of the risk that is uniformly consistent over the spaces could replace the exact leave-one-out risk, and one with smaller fluctuation may reduce the estimation error that each dimension adds to \eqref{eq:krylov-excess}.

Several questions remain open. The practical choices of $m$ and $K$ have no guarantees yet. The schedule of \cref{thm:sieve} is the only data-independent choice we analyze, and its rate is likely improvable by bounding Chebyshev polynomials on the real spectrum rather than on a complex contour. In the simulations of \cref{sec:sim-sieve} the estimation cost per dimension did not grow with $m$, and the best dimension exceeded the schedule. 
\cref{thm:consistency} requires a polynomial tail bound on $\|S_n\|_\op$, which excludes elliptical designs with heavy-tailed radius. Exact leave-one-out costs one $n\times n$ eigendecomposition, and this cost is no longer negligible for very large $n$. \Cref{thm:local} needs only the local response moments $(T_n^ay)_i$, which cost $m$ matrix-vector products, and the diagonals $(T_n^j)_{ii}$. Replacing these diagonals by their average, which for ridge regression gives generalized cross-validation, fails under elliptical designs (\cref{rem:gcv}), so combining \cref{thm:local} with randomized estimation of the diagonals is a natural direction.

\section*{Declaration of the use of generative AI and AI-assisted technologies}
During the preparation of this work the authors used Claude and Claude Code (Anthropic) in order to condense the manuscript, edit its structure and wording, check the theorems and their proofs, and organize the code. After using these tools the authors reviewed and edited the content as necessary and take full responsibility for the content of the publication.

\section*{Supplementary material}\label{SM}
The Supplementary Material includes the proofs of all results not proved in the main text or the Appendix, implementation details, additional simulation results and details of the real-data applications. The implementation details cover the computation of the features of the separated directions and of leave-one-out predictions along gradient descent paths, and the Chebyshev form of \cref{alg:loo} used in our computations. The additional simulation results include a comparison of rules for the truncation.
Python code that reproduces the simulations of \cref{sec:sims} is available at \url{https://doi.org/10.5281/zenodo.23755357}.

\appendix

\appendixone
\section*{Appendix 1}
\subsection*{Proofs of \cref{thm:local} and \cref{prop:trunc-loo}}\phantomsection\label{app:proofs}

\begin{lemma}\label{lem:schur}
Let $A\in\R^{n\times n}$ be symmetric, let $A^{(i)}$ be $A$ without its $i$-th row and column and $\alpha_i=A_{-i,i}\in\R^{n-1}$ its $i$-th column without the diagonal entry, and let $z\in\Cx$ with $z\notin\spec(A)\cup\spec(A^{(i)})$. Write $D=zI-A^{(i)}$ and $\omega_i(z)=y_i+\alpha_i^\top D^{-1}y^{(i)}$. Then $((zI-A)^{-1})_{ii}=(z-A_{ii}-\alpha_i^\top D^{-1}\alpha_i)^{-1}\neq0$ and
\[
((zI-A)^{-1}y)_i=((zI-A)^{-1})_{ii}\,\omega_i(z).
\]
\end{lemma}

\begin{proof}
Order the coordinate $i$ first, so that $zI-A$ has blocks $z-A_{ii}$, $-\alpha_i^\top$, $-\alpha_i$ and $D$, and let $\tau=z-A_{ii}-\alpha_i^\top D^{-1}\alpha_i$. Since $\det(zI-A)=\det(D)\tau$ and $zI-A$ is invertible, $\tau\neq0$. Block inversion gives
\[
(zI-A)^{-1}=\begin{pmatrix}\tau^{-1}&\tau^{-1}\alpha_i^\top D^{-1}\\ \tau^{-1}D^{-1}\alpha_i \quad &D^{-1}+\tau^{-1}D^{-1}\alpha_i\alpha_i^\top D^{-1}\end{pmatrix}.
\]
The first row applied to $y$ gives $((zI-A)^{-1}y)_i=\tau^{-1}\omega_i(z)$.
\end{proof}

\begin{proof}[of \cref{thm:local}]
We prove \eqref{eq:local-rec} for a general symmetric $A$ in place of $T_n$, with $v_{i,a}=(A^ay)_i$ and $u_{i,a}=\alpha_i^\top(A^{(i)})^{a-1}y^{(i)}$ for $a\geq1$ and $u_{i,0}=v_{i,0}=y_i$. For $|z|>\|A\|_\op$, which also exceeds $\|A^{(i)}\|_\op$ by interlacing, the Neumann series converge absolutely and give
\[
((zI-A)^{-1})_{ii}=\sum_{j\geq0}\frac{(A^j)_{ii}}{z^{j+1}},\quad((zI-A)^{-1}y)_i=\sum_{a\geq0}\frac{v_{i,a}}{z^{a+1}},\quad\omega_i(z)=\sum_{a\geq0}\frac{u_{i,a}}{z^{a}},
\]
by the definition of $u_{i,a}$. By \cref{lem:schur}, the second series is the product of the first and the third. Comparing the coefficients of $z^{-a-1}$ gives $v_{i,a}=\sum_{j=0}^a(A^j)_{ii}u_{i,a-j}$, which is \eqref{eq:local-rec} because $(A^0)_{ii}=1$.

For $A=T_n$ we have $\alpha_i=t_i=X^{(i)}x_i/n$, where $X^{(i)}$ is $X$ without its $i$-th row. Since $S_n^{(i)}X^{(i)\top}=X^{(i)\top}T_n^{(i)}$, induction gives $(S_n^{(i)})^aX^{(i)\top}=X^{(i)\top}(T_n^{(i)})^a$, and therefore $x_i^\top(S_n^{(i)})^as_n^{(i)}=x_i^\top X^{(i)\top}(T_n^{(i)})^ay^{(i)}/n=t_i^\top(T_n^{(i)})^ay^{(i)}$. The leave-one-out moments of \cref{thm:local} are thus the $u_{i,a}$ above with $A=T_n$.
\end{proof}

\begin{proof}[of \cref{prop:trunc-loo}]
Let $X^{(i)}=\sum_r\sigma_ra_rb_r^\top$ be a singular value decomposition with $\sigma_r>0$. Then $S_n^{(i)}b_r=(\sigma_r^2/n)b_r$, $T_n^{(i)}a_r=(\sigma_r^2/n)a_r$, $s_n^{(i)}=n^{-1}\sum_r\sigma_rb_r(a_r^\top y^{(i)})$, and $t_i=X^{(i)}x_i/n$ is orthogonal to the null space of $T_n^{(i)}$. Hence, for every function $\phi$ on the spectrum, $x_i^\top\phi(S_n^{(i)})s_n^{(i)}=t_i^\top\phi(T_n^{(i)})y^{(i)}$. Let $\mu_1>\mu_2>\cdots$ be the distinct eigenvalues of $T_n^{(i)}$, $\Pi_r$ the corresponding spectral projectors and $\kappa_r=t_i^\top\Pi_ry^{(i)}$, so that $t_i^\top\phi(T_n^{(i)})y^{(i)}=\sum_r\phi(\mu_r)\kappa_r$. By \cref{lem:schur} with $A=T_n$ and the eigendecomposition of $T_n$, for $z$ outside both spectra,
\begin{equation}\label{eq:partial-fractions}
\frac{N_i(z)}{D_i(z)}-y_i=\sum_r\frac{\kappa_r}{z-\mu_r},\qquad N_i(z)=\sum_l\frac{Q_{il}\tilde y_l}{z-\hat\lambda_l},\quad D_i(z)=\sum_l\frac{Q_{il}^2}{z-\hat\lambda_l}=-\psi_i(z).
\end{equation}
On $(\hat\lambda_{j+1},\hat\lambda_j)$, $\psi_i'(\theta)=\sum_lQ_{il}^2/(\hat\lambda_l-\theta)^2>0$, and $\psi_i(\theta)\to+\infty$ as $\theta\uparrow \hat\lambda_j$ and $\psi_i(\theta)\to-\infty$ as $\theta\downarrow \hat\lambda_{j+1}$, because $Q_{ij}$ and $Q_{i,j+1}$ are nonzero. Hence $\psi_i$ has exactly one root $\hat\lambda^{(i)}_j$ there. With $e_i$ the $i$-th standard basis vector of $\R^n$, the vector $\tilde q=(T_n-\hat\lambda^{(i)}_jI)^{-1}e_i$ has $i$-th coordinate $\psi_i(\hat\lambda^{(i)}_j)=0$, so the rows $k\neq i$ of $(T_n-\hat\lambda^{(i)}_jI)\tilde q=e_i$ give $(T_n^{(i)}-\hat\lambda^{(i)}_jI)\tilde q^{(i)}=0$ with $\tilde q^{(i)}\neq0$, and $\hat\lambda^{(i)}_j$ is an eigenvalue of $T_n^{(i)}$. Cauchy interlacing places the $j$-th largest eigenvalue of $T_n^{(i)}$, counted with multiplicity, in $[\hat\lambda_{j+1},\hat\lambda_j]$ and the neighboring ones in $[\hat\lambda_j,\hat\lambda_{j-1}]$ and $[\hat\lambda_{j+2},\hat\lambda_{j+1}]$, with $\hat\lambda_0=\infty$. Since the top $K+1$ eigenvalues of $T_n$ are simple, $\hat\lambda^{(i)}_j$ is the $j$-th largest eigenvalue $\mu_j$ of $T_n^{(i)}$ and is simple. As $\hat\lambda^{(i)}_j$ is a simple zero of $D_i$, with $D_i'(\hat\lambda^{(i)}_j)=-\sum_lQ_{il}^2/(\hat\lambda_l-\hat\lambda^{(i)}_j)^2<0$, and not a pole of $N_i$, the residue of the left-hand side of \eqref{eq:partial-fractions} at $\hat\lambda^{(i)}_j$ is $N_i(\hat\lambda^{(i)}_j)/D_i'(\hat\lambda^{(i)}_j)=o_{ij}$, and that of the right-hand side is $\kappa_j$. The $j$-th separated direction corresponds to $\phi=1\{\lambda=\mu_j\}$, whose feature is $\kappa_j=o_{ij}$, and the $k$-th bulk basis vector to $\phi(\lambda)=\tilde\varphi_k(\lambda)1\{\lambda\notin\{\mu_1,\ldots,\mu_K\}\}$, whose feature is $\sum_r\tilde\varphi_k(\mu_r)\kappa_r-\sum_{j\leq K}\tilde\varphi_k(\mu_j)\kappa_j=t_i^\top\tilde\varphi_k(T_n^{(i)})y^{(i)}-\sum_{j\leq K}\tilde\varphi_k(\hat\lambda^{(i)}_j)o_{ij}$.
\end{proof}

\bibliographystyle{biometrika}
\bibliography{references}

\end{document}